\documentstyle[twoside,amstex,epsfig,amssymb,here]{article}

\catcode`\@=11
\long\def\@makefntext#1{
\protect\noindent \hbox to 3.2pt {\hskip-.9pt  
$^{{\eightrm\@thefnmark}}$\hfil}#1\hfill}		

\def\@makefnmark{\hbox to 0pt{$^{\@thefnmark}$\hss}}	
	
\def\ps@myheadings{\let\@mkboth\@gobbletwo
\def\@oddhead{\hbox{}
\rightmark\hfil\eightrm\thepage}   
\def\@oddfoot{}\def\@evenhead{\eightrm\thepage\hfil
\leftmark\hbox{}}\def\@evenfoot{}
\def\sectionmark##1{}\def\subsectionmark##1{}}

\oddsidemargin=\evensidemargin
\newcounter{sectionc}\newcounter{subsectionc}\newcounter{subsubsectionc}
\renewcommand{\section}[1] {\vspace{12pt}\addtocounter{sectionc}{1} 
\setcounter{subsectionc}{0}\setcounter{subsubsectionc}{0}\noindent 
	{\tenbf\thesectionc. #1}\par\vspace{5pt}}
\renewcommand{\subsection}[1] {\vspace{12pt}\addtocounter{subsectionc}{1} 
	\setcounter{subsubsectionc}{0}\noindent 
	{\bf\thesectionc.\thesubsectionc. {\kern1pt \bfit #1}}\par\vspace{5pt}}
\renewcommand{\subsubsection}[1] {\vspace{12pt}\addtocounter{subsubsectionc}{1}
	\noindent{\tenrm\thesectionc.\thesubsectionc.\thesubsubsectionc.
	{\kern1pt \tenit #1}}\par\vspace{5pt}}
\newcommand{\nonumsection}[1] {\vspace{12pt}\noindent{\tenbf #1}
	\par\vspace{5pt}}

\newcounter{appendixc}
\newcounter{subappendixc}[appendixc]
\newcounter{subsubappendixc}[subappendixc]
\renewcommand{\thesubappendixc}{\Alph{appendixc}.\arabic{subappendixc}}
\renewcommand{\thesubsubappendixc}
	{\Alph{appendixc}.\arabic{subappendixc}.\arabic{subsubappendixc}}

\renewcommand{\appendix}[1] {\vspace{12pt}
        \refstepcounter{appendixc}
        \setcounter{figure}{0}
        \setcounter{table}{0}
        \setcounter{lemma}{0}
        \setcounter{theorem}{0}
        \setcounter{corollary}{0}
        \setcounter{definition}{0}
        \setcounter{equation}{0}
        \renewcommand{\thefigure}{\Alph{appendixc}.\arabic{figure}}
        \renewcommand{\thetable}{\Alph{appendixc}.\arabic{table}}
        \renewcommand{\theappendixc}{\Alph{appendixc}}
        \renewcommand{\thelemma}{\Alph{appendixc}.\arabic{lemma}}
        \renewcommand{\thetheorem}{\Alph{appendixc}.\arabic{theorem}}
        \renewcommand{\thedefinition}{\Alph{appendixc}.\arabic{definition}}
        \renewcommand{\thecorollary}{\Alph{appendixc}.\arabic{corollary}}
        \renewcommand{\theequation}{\Alph{appendixc}.\arabic{equation}}
        \noindent{\tenbf Appendix \theappendixc #1}\par\vspace{5pt}}
\newcommand{\subappendix}[1] {\vspace{12pt}
        \refstepcounter{subappendixc}
        \noindent{\bf Appendix \thesubappendixc. {\kern1pt \bfit #1}}
	\par\vspace{5pt}}
\newcommand{\subsubappendix}[1] {\vspace{12pt}
        \refstepcounter{subsubappendixc}
        \noindent{\rm Appendix \thesubsubappendixc. {\kern1pt \tenit #1}}
	\par\vspace{5pt}}

\newcommand{\textlineskip}{\baselineskip=13pt}
\newcommand{\smalllineskip}{\baselineskip=10pt}

\def\eightcirc{
\begin{picture}(0,0)
\put(4.4,1.8){\circle{6.5}}
\end{picture}}
\def\eightcopyright{\eightcirc\kern2.7pt\hbox{\eightrm c}} 

\newcommand{\copyrightheading}[1]
	{\vspace*{-2.5cm}\smalllineskip{\flushleft
	{\footnotesize International Journal of Modern Physics A, #1}\\
	{\footnotesize $\eightcopyright$\, World Scientific Publishing
	 Company}\\
	 }}

\def\abstracts#1#2#3{{
	\centering{\begin{minipage}{4.5in}\baselineskip=10pt\footnotesize
	\parindent=0pt #1\par 
	\parindent=15pt #2\par
	\parindent=15pt #3
	\end{minipage}}\par}} 

\renewenvironment{thebibliography}[1]
	{\frenchspacing
	 \ninerm\baselineskip=11pt
	 \begin{list}{\arabic{enumi}.}
	{\usecounter{enumi}\setlength{\parsep}{0pt}
	 \setlength{\leftmargin 12.7pt}{\rightmargin 0pt} 
	 \setlength{\itemsep}{0pt} \settowidth
	{\labelwidth}{#1.}\sloppy}}{\end{list}}

\newcounter{itemlistc}
\newcounter{romanlistc}
\newcounter{alphlistc}
\newcounter{arabiclistc}

\newcommand{\fcaption}[1]{
        \refstepcounter{figure}
        \setbox\@tempboxa = \hbox{\footnotesize Fig.~\thefigure. #1}
        \ifdim \wd\@tempboxa > 5in
           {\begin{center}
        \parbox{5in}{\footnotesize\smalllineskip Fig.~\thefigure. #1}
            \end{center}}
        \else
             {\begin{center}
             {\footnotesize Fig.~\thefigure. #1}
              \end{center}}
        \fi}

\newcommand{\tcaption}[1]{
        \refstepcounter{table}
        \setbox\@tempboxa = \hbox{\footnotesize Table~\thetable. #1}
        \ifdim \wd\@tempboxa > 5in
           {\begin{center}
        \parbox{5in}{\footnotesize\smalllineskip Table~\thetable. #1}
            \end{center}}
        \else
             {\begin{center}
             {\footnotesize Table~\thetable. #1}
              \end{center}}
        \fi}

\def\@citex[#1]#2{\if@filesw\immediate\write\@auxout
	{\string\citation{#2}}\fi
\def\@citea{}\@cite{\@for\@citeb:=#2\do
	{\@citea\def\@citea{,}\@ifundefined
	{b@\@citeb}{{\bf ?}\@warning
	{Citation `\@citeb' on page \thepage \space undefined}}
	{\csname b@\@citeb\endcsname}}}{#1}}

\newif\if@cghi
\def\cite{\@cghitrue\@ifnextchar [{\@tempswatrue
	\@citex}{\@tempswafalse\@citex[]}}
\def\citelow{\@cghifalse\@ifnextchar [{\@tempswatrue
	\@citex}{\@tempswafalse\@citex[]}}
\def\@cite#1#2{{$\null^{#1}$\if@tempswa\typeout
	{IJCGA warning: optional citation argument 
	ignored: `#2'} \fi}}

\def\pmb#1{\setbox0=\hbox{#1}
	\kern-.025em\copy0\kern-\wd0
	\kern.05em\copy0\kern-\wd0
	\kern-.025em\raise.0433em\box0}

\def\fnt#1#2{\footnotetext{\kern-.3em
	{$^{\mbox{\scriptsize #1}}$}{#2}}}

\def\fpage#1{\begingroup
\voffset=.3in
\thispagestyle{empty}\begin{table}[b]\centerline{\footnotesize #1}
	\end{table}\endgroup}

\def\runninghead#1#2{\pagestyle{myheadings}
\markboth{{\protect\footnotesize\it{\quad #1}}\hfill}
{\hfill{\protect\footnotesize\it{#2\quad}}}}
\font\tenrm=cmr10
\font\tenit=cmti10 
\font\tenbf=cmbx10
\font\bfit=cmbxti10 at 10pt
\font\ninerm=cmr9

\font\eightrm=cmr8

\def\qed{\hbox{${\vcenter{\vbox{			
   \hrule height 0.4pt\hbox{\vrule width 0.4pt height 6pt
   \kern5pt\vrule width 0.4pt}\hrule height 0.4pt}}}$}}

\begin{document}

\runninghead{Physics of W bosons at LEP2} {Physics of W bosons at LEP2}

\normalsize\textlineskip
\thispagestyle{empty}
\setcounter{page}{1}

\copyrightheading{}			

\vspace*{0.88truein}

\fpage{1}
\centerline{\bf Physics of W bosons at LEP2}
\vspace*{0.035truein}

\vspace*{0.37truein}
\centerline{\footnotesize Mario Campanelli}
\vspace*{0.015truein}
\centerline{\footnotesize\it Institut f\"ur Teilchenphysik}
\baselineskip=10pt
\centerline{\footnotesize\it ETH H\"onggerberg CH-8093 Z\"urich}
\vspace*{10pt}
\begin{center}
To be published in International Journal of Modern Physics A
\end{center}

\vspace*{0.21truein}
\abstracts{After the first observations of W bosons in leptonic interactions,
about 4000 WW candidate events per experiment have been collected at LEP2.
This data allows the measurement of the WW production cross section at
different centre-of-mass energies, as well as W decay branching fractions.
The W hadronic branching fraction can be converted into a test of the
unitarity of the CKM matrix, or into an indirect determination of the
matrix element $|V_{cs}|$. A more direct measurement coming from charm
tagging is also performed.
The W mass has been measured via the cross section (in the threshold region)
and the direct reconstruction of the W decay products, using different
techniques to account for the distortions due to experimental effects.
The main systematic error to the mass reconstruction in the fully hadronic
channel comes from QCD effects like Color reconnections and Bose-Einstein 
correlations, extensively studied in WW events.
In $e^+e^-$ collisions W pairs can be produced in s-channel via a three
vector boson vertex, so a direct study of the trilinear gauge boson couplings
is possible. Modification of WW cross section and distributions of W 
production and decay angles would be an indication of non-standard couplings,
thus a first hint for the presence of new physics.
}{}{}


\vspace*{1pt}\textlineskip	
\section{Introduction}
The experimental program of the LEP accelerator at CERN was foreseen to 
proceed in two steps: a first period of running around the energy of the Z 
boson, and a
second period at higher energy, having as main goals the production of W 
boson pairs and the search for new particles.\par
W bosons can be studied at LEP in a unique environment. Fundamental ingredients
of the Standard Model\cite{sem} as carriers of the charged electroweak 
interaction, these particles were discovered in 1983 in $p\bar{p}$ collisions
by the UA1 and UA2 collaborations at CERN\cite{ua1},\cite{ua2},\cite{wzdisc}. 
Further,
more precise measurements were performed by the CDF and D0 experiments running
at the Tevatron collider at Fermilab \cite{cdfd0}.\par 
At LEP it is therefore the
first time that W bosons are produced in the clean environment of leptonic
interactions. In the hadronic case the most copious source
of Ws is the Drell-Yan mechanism, with production and subsequent decay of 
single Ws. Due to the large QCD background to the hadronic decay channel, 
most of the measurements performed are relative to the cleaner decay channels
$W\to e\bar{\nu_e}$ and $W\to \mu\bar{\nu_\mu}$. In $e^+e^-$ interactions,
W bosons are mainly produced in pairs, and according to their decay WW
events can be classified as fully leptonic, semileptonic and fully hadronic.
Above the WW production threshold, all decay channels can be studied with 
small background contamination, giving a broader picture of the physics of 
these particles.\par
Measurements of WW and single W production cross sections can be performed,
as well as W decay branching ratios, providing a test of lepton universality
for charged current interactions.\par 
As it will be discussed in more detail in the next sections, the W mass is
one of the fundamental parameters of the Standard Model. Its actual value
depends via radiative corrections from unknown parameters like the mass of the
Higgs boson, or on the presence of physics beyond the Standard Model.
The error on this quantity from the measurements performed at LEP2 is 
presently the same as that coming from hadronic interactions, and
being still dominated by statistics, it will improve in the next years of
data taking. Preliminary studies \cite{yrep} have shown that with the target 
luminosity of 500 $pb^{-1}$ LEP2 can reach a precision on this quantity 
$\Delta M_W=50$ MeV, with a factor 2 improvement with respect to the present 
measurements.\par
The main limitations to the accuracy achievable on the mass are coming from
the LEP energy measurement and to final state interactions leading to a 
distortion of the reconstructed W mass. Since these effect can produce
sensible mass shifts, as well as modification in other observables, several 
models have been proposed and tested with the available data.\\
The full reconstruction of final states is extremely important for the study of
Trilinear Gauge Couplings. S-channel production of W bosons occurs via diagrams
involving $\gamma WW$ and $ZWW$ vertices. A deviation from the standard model
for these vertices can modify WW production cross section, as well as
distributions of W production and decay angles. Constraints on anomalous
couplings are coming from the combined studies of WW events, as well as
single-W and single-$\gamma$.\par
If standard model couplings are assumed, W hadronic branching fractions are 
proportional to squares of CKM matrix elements $|V_{ab}|^2$. The matrix element
$|V_{cs}|$ is in particular presently known with worse precision with respect
to the others; assuming the unitarity of the matrix and the knowledge of the
other matrix elements, a more precise determination of this quantity can be 
derived from the WW hadronic branching fraction. A less precise but more 
direct determination can be obtained from a charm tagging of the jets from
W decays, exploiting heavy-quark characteristics of charm in an environment
with small contamination from b quarks.

\section{Tree-level relations for gauge bosons}
The unified theory of weak and electromagnetic interaction is based
on the invariance of the Lagrangian under transformations of the 
$SU(2)_L \times U(1)$ symmetry group. The three fields $(W^1_\mu,W^2_\mu,W^3_\mu)$
are connected to the weak isospin T, and the field $B_\mu$ to
the weak hypercharge Y. These quantum numbers are related to the electric 
charge Q by $Q=T_3+Y/2$, where $T_3$ is the third component of the
weak isospin. The physical fields, the carriers of the charged (W$^\pm$) and
neutral (Z) weak currents, and of the electromagnetic current (A) are
linear combinations of the above
\[W^\pm_\mu=\frac{1}{\sqrt{2}}(W^1_\mu\mp iW^2_\mu)\]
\[\binom{A_\mu}{Z_\mu}=\begin{pmatrix}\cos\theta_W&\sin\theta_W\\
-\sin\theta_W&\cos\theta_W\end{pmatrix}\binom{B_\mu}{W^0_\mu}\]
The weak angle $\sin\theta_W$ introduced relates the coupling constants 
of the $SU(2)_L$ and $U(1)$ interactions to the electric charge
\[g=\frac{e}{\sin\theta_W}\]
\[g'=\frac{e}{\cos\theta_W}\]
At low energies, the electroweak theory is equivalent to the Fermi
theory of the weak interactions. The Fermi constant can be expressed as
\[ G_F=\frac{g^2}{4\sqrt{2}M_W^2}=\frac{\pi\alpha_{QED}}{\sqrt{2}M_W^2
\sin^2\theta_W}\]
Since vector bosons behave as real particles, a gauge-invariant kinetic term
must be added to the Lagrangian describing electroweak interactions.
Due to the non-Abelian structure of the gauge field, the commutators of the
covariant derivatives involved do not vanish, but produce terms leading to
the self-interaction of the gauge bosons:
\[{\cal L}_W^{kin}=-\frac{1}{4}W^j_{\mu\nu}W_j^{\mu\nu}\]
\[W^j_{\mu\nu}=\partial_\mu W^j_\nu-\partial_\nu W^j_\mu+
g\epsilon^j_{km}W^k_\mu W^m_\nu\]
The last term involves the product of two W fields, so in the Lagrangian 
terms
for trilinear and quadrilinear gauge boson interactions are present.\par
In the standard model the vector bosons acquire a mass by the spontaneous
breaking of the symmetry group, introducing a Higgs \cite{higgs} doublet
\[\binom{\Phi^+}{\Phi^0}=\binom{\Phi_1+i\Phi_2}{\Phi_3+i\Phi_4} \]
and choosing a vacuum expectation value $|\Phi^0|^2=v^2/2>0$. 
The masses of the vector
bosons are then determined by this vacuum expectation value and the coupling
constants:
\[m_W=\frac{gv}{2},\hspace{2cm}m_Z=\frac{\sqrt{g^2+{g'}^2}}{2}v,\hspace{2cm}
m_\gamma=0\]
which leads to a relation between the masses of the vector bosons and the
weak angle
\[\cos^2\theta_W=\frac{m^2_W}{\rho m^2_Z}\]
The $\rho$ parameter is equal to unity at tree level. Deviations from
this value can arise from radiative corrections.\par
\section{Indirect determination of the W mass}
\begin{figure}[tb]
  \begin{center}
   \includegraphics[width=8cm]{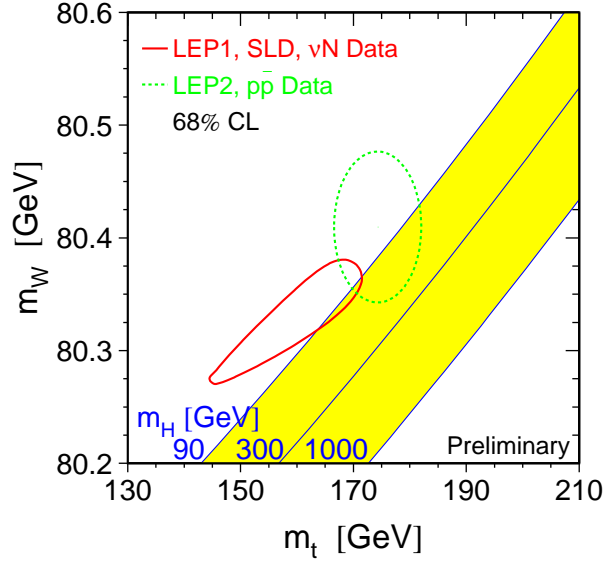}
  \end{center}
  \caption{Comparison between the direct measurements of the masses of W 
and top mass and the predictions from the electroweak fit, for various
values of the Higgs mass} \label{fig:mwmtew99}
\end{figure}

To obtain accurate predictions, the tree-level relations presented above
are no longer sufficient, but the evaluation of higher orders is needed,
especially given the accuracy of the available data. 
 An example is the possibility of extracting the W
mass from the standard model relations without directly measuring this 
quantity.
From the equations shown in the previous section, it is possible to derive
\[\sqrt{2}G_FM^2_W(1-\frac{M^2_W}{\rho M^2_Z})=\pi\alpha\]
To include higher-order effects, different schemes can be used. The most 
common approach \cite{ybooklep1} is to keep $\rho$ to its tree level 
value 1 and include
all corrections in a quantity $\Delta r$, that accounts for both weak
and electromagnetic effects:
\[\sqrt{2}G_FM^2_W(1-\frac{M^2_W}{M^2_Z})=\frac{\pi\alpha}{1-\Delta r}\]
Since $\alpha$, $G_F$ and $M_Z$ are experimentally well known 
quantities, it is possible to derive the W mass from
\[M_W^2=\frac{1}{2}M_Z^2(1+\sqrt{1-\frac{4A_0^2}{M_Z^2}\frac{1}{1-\Delta r}})\]
with
\[A_0^2=\frac{\pi\alpha}{\sqrt{2}G_F}\]
In the standard model, vertex and propagator corrections can be 
decoupled into an
electromagnetic part, due to the running of the coupling constant 
$\alpha_{QED}$, and a weak part, that contains terms showing a quadratic
dependence on the top mass and a logarithmic dependence on the Higgs mass:
\[\Delta r=\Delta\alpha+\frac{\cos^2\theta_W}{\sin^2\theta_W}\frac{3 G_\mu
m_t^2}{8\pi^2\sqrt{2}}+\frac{\sqrt{2}G_\mu M_W^2}{16\pi^2}[\frac{11}{3}
(\log\frac{m_H^2}{M_W^2}-\frac{5}{6})]+... \]
The mass of those particles enters therefore as a parameter to the indirect
determination of the W mass, as can be seen in figure \ref{fig:mwmtew99}
(full curve), where a clear correlation emerges between the indirect 
determinations of the masses of the W and the top quark, as extracted from 
a fit to precision electroweak data (mainly coming from LEP1 measurements) 
available in winter 1999 \cite{ewmeas99}.\par
The plot also shows as a dotted ellipse the direct measurement of both
masses, in good agreement with the indirect predictions. It is possible then
to also include the measured value of $m_t$ and $m_W$ in the fit, and get 
some indication on the mass of the Higgs boson.\par
It is therefore very important to improve the precision on the direct 
determination of the W mass, to further constraint the standard model, and
get stronger bounds on the allowed range for the Higgs mass.\par
\section{Models for Trilinear Gauge Coupling}
We have already seen that in the standard model vertices involving gauge
bosons only derive from the request of gauge invariance of the kinetic term.
Since trilinear couplings are extensively studied at LEP2, possible
deviations from the standard model value will be discussed.\par
The most general Lorenz-invariant effective Lagrangian, expressing the 
coupling of two oppositely charged 
and one neutral vector bosons is the following\cite{ybooktgc}:
\[ {\cal L}_{WWV}^{eff}/g_{WWV}=ig_1^V(W_{\mu\nu}^\dagger W^\mu V^\nu-
W_\mu^\dagger V_\nu W^{\mu\nu})+i k_V W_\mu^\dagger W_\nu V^{\mu\nu}\]
\[+\frac{i\lambda_V}{m_W^2}W_{\mu\nu}^\dagger W^\mu_\rho V^{\nu\rho}-
g_4^V W_\mu^\dagger W_\nu(\partial^\mu V^\nu+\partial^\nu V^\mu)\]
\[+g_5^V\epsilon^{\mu\nu\rho\sigma}(W^\dagger_\mu \overrightarrow{\partial_\rho}
W_\nu)V_\sigma+i \tilde{k}_V W_\mu^\dagger W_\nu\tilde{V}^{\mu\nu}\]
\[+\frac{i\tilde{\lambda}_V}{m_W^2}W_{\mu\nu}^\dagger W_\rho^\mu \tilde{V}^{\nu\rho}\]
Here V stands for either a photon or a Z boson ($V=\gamma,Z$), and W for the
W field. $g_{WWV}$ are fixed to
\[ g_{WW\gamma}=-e   \hspace{3cm}  g_{WWZ}=-e \cot \theta_W.\]
At tree level, the SM predicts $g_1^Z=g_1^\gamma=k_Z=k_\gamma=1$, with all
other couplings vanishing. The terms $g_1^V, k_V$ and
$\lambda_V$ conserve C and P separately, while $g_5^V$ violates both C and P
conserving CP. The coupling between W and photons can be related to intuitive
physical quantities; in particular the terms conserving C and P correspond to 
the lowest-order terms in a multipole expansion of the interactions between Ws
and photons; thus they can be related to the magnetic moment $\mu_W$ and the
electric quadrupole moment $Q_W$:
\[\mu_W=\frac{e}{2 m_W}(1+k_\gamma+\lambda_\gamma)\]
\[Q_W=-\frac{e}{m_W^2}(k_\gamma-\lambda_\gamma)\]
The two parity-violating couplings $\tilde{k}_\gamma$ and $\tilde{\lambda}
_\gamma$ respect charge-conjugation invariance, and are related to the electric
dipole moment $d_W$ and to the magnetic quadrupole moment $\tilde{Q}_W$:
\[d_W=\frac{e}{2m_W}(\tilde{k}_\gamma+\tilde{\lambda}_\gamma)\]
\[\tilde{Q}_W=-\frac{e}{m_W^2}(\tilde{k}_\gamma-\tilde{\lambda}_\gamma).\]
Due to the relatively limited statistics available in the present experimental
facilities, the set of free parameters present in the general Lagrangian
quoted above is too large for practical uses. This set of parameters can be
reduced under a certain number of assumptions, depending on the way the
effective Lagrangian quoted above is made gauge-invariant, i.e. what kind of 
new physics is expected to generate the couplings. If a light Higgs boson is
present, and considering only the C- and P- conserving operators, the effective
Lagrangian can take the gauge-invariant form
\[{\cal L}^{TGC}=ig'\frac{\alpha_{B\Phi}}{m_W^2}(D_\mu \Phi)^\dagger B^{\mu\nu}
(D_\nu \Phi)+ig\frac{\alpha_{W\Phi}}{m_W^2}(D_\mu \Phi)^\dagger \tau W^{\mu\nu}
(D-\nu\Phi)+g\frac{\alpha_W}{6m_W^2}W^\mu_\nu(W^\nu_\rho\times W^\rho_\mu)\]
with g and g' the SM couplings of the $SU(2)_L$ and $U(1)_Y$ symmetries.
When the Higgs field is replaced by its vacuum expectation value 
$(0,v/\sqrt{2})$, the following relations can be written:
\[\Delta g^Z_1=g^Z-1=\frac{\alpha_{W\Phi}}{\cos^2\theta_W}\]
\[\Delta k_\gamma=(k_\gamma-1)=-\frac{\cos^2\theta_W}{\sin^2\theta_W}
(\Delta k_Z-\Delta g_1^Z)=\alpha_{W\Phi}+\alpha_{B\Phi}\]
\[\lambda_\gamma=\lambda_Z=\alpha_W\]
and is then natural to use the above relations and express all measured
quantities as a function of $(\Delta g_1^Z,\Delta k_\gamma,\lambda_\gamma)$ or the $\alpha$ parameters.
\par
In the absence of a light Higgs, a non linear approach can be followed to 
make the effective Lagrangian gauge-invariant. In this scheme, it is convenient
to express the couplings as a function of the lowest-dimension operators 
$(\Delta g_1^Z,\Delta k_\gamma,\Delta k_Z)$, 
while the parameters $\lambda_\gamma$ and $\lambda_Z$ are usually set to zero.
\par
\section{Four-fermion production in $e^+e^-$ collisions}
Processes involving W pair production in e$^+$e$^-$ collisions are a subset
of a larger set of diagrams contributing to four-fermion final state and
interfering with each other, so all of them have to be considered when dealing 
with W events. Processes contributing to 4-fermion final states can be 
divided into charged current (CC) and neutral current (NC). The first class
comprises production of (up, antidown) and (down-antiup)-type fermion pairs,
where each pair has the same generation index. Final states produced via
W production belong to this class. Neutral current events are those where
two fermion-antifermion pairs are produced, and they
are mediated by the neutral gauge bosons. Obviously these two classes
overlap for certain final states. The number of Feynman diagrams 
contributing to the charged current class is shown in table \ref{tab:cc}
for the possible combinations of final states. Three different cases occur
(shown in the table by different character types):
\begin{itemize}
\item {\bf CC11 family (boldface)}: for two different fermion pairs, none
of which is an electron, or electron neutrino, no identical particles are
involved, and there is at maximum 11 diagrams.
\item CC20 family (normal): one e$^\pm$ and one $\nu_e$ are in the final 
state, so additional diagrams with t-channel exchange of the gauge boson are
present
\item {\it CC43/mix43 CC56/mix56 (italics)}: two mutually charge 
conjugate pairs are produced, so these diagrams can proceed via both 
charged and neutral gauge boson exchange
\end{itemize}
\begin{table}\begin{center}
\begin{tabular}{|c|c|c|c|c|c|}\hline
&$\bar{d}u$&$\bar{s}c$&$\bar{e}\nu_e$&$\bar{\mu}\nu_\mu$&$\bar{\tau}\nu_\tau$\\ \hline
$d\bar{u}$&{\it 43}&{\bf 11}&20&{\bf 10}&{\bf 10}\\
$e\bar{\nu_e}$&20&20&{\it 56}&18&18\\
$\mu\bar{\nu_\mu}$&{\bf 10}&{\bf 10}&18&{\it 19}&{\bf 9}\\ \hline
\end{tabular}
\caption{Number of diagrams for Charged Current final states}
\label{tab:cc}
\end{center}\end{table}
\par
\begin{figure}[p]
  \begin{center}
   \includegraphics[width=14cm,bbllx=40,bblly=100,bburx=550,bbury=685]{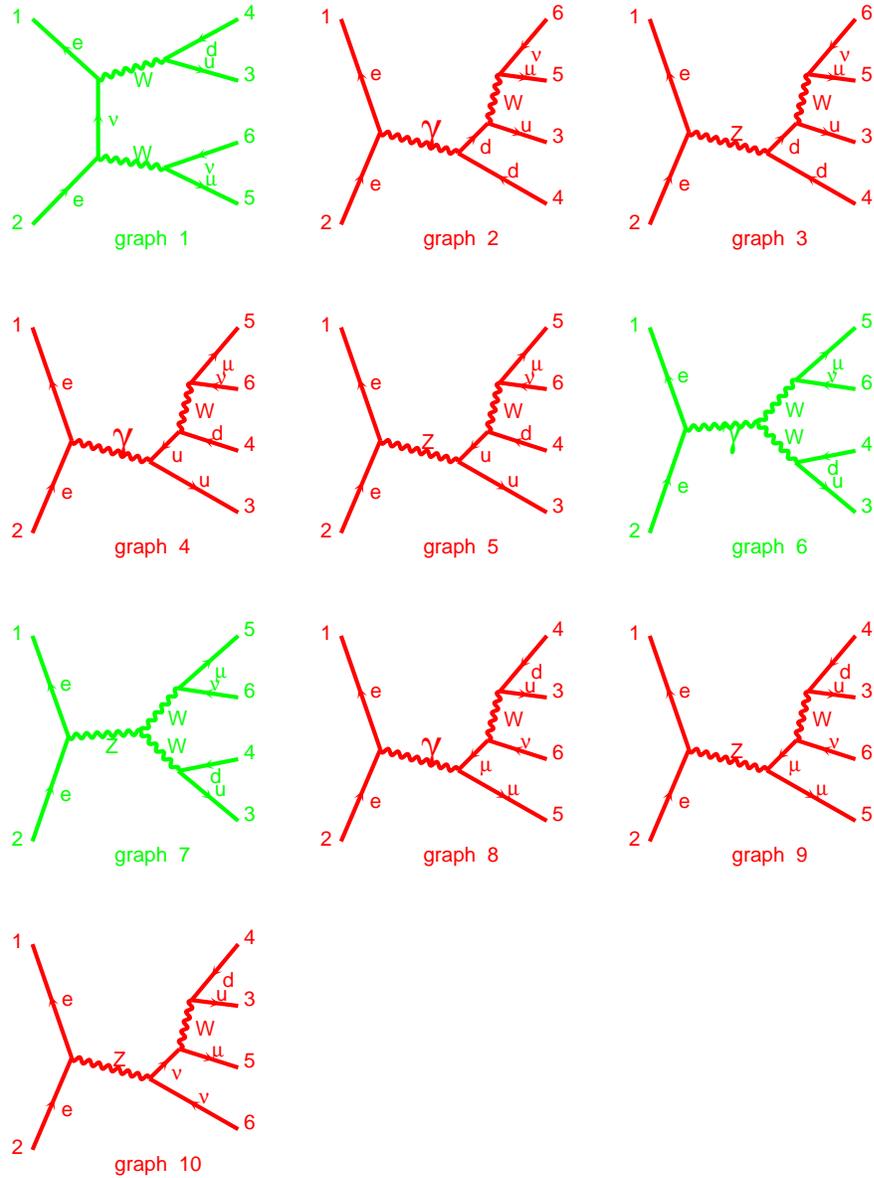}
  \end{center}
  \caption{Feynman diagrams involved in $e^+e^-\to u\bar{d}
\mu\bar{\nu_\mu}$ final 
state. Graphs in light grey (1, 6 and 7) correspond to W pair production 
(CC03)} \label{fig:feyncc11}
\end{figure}
\par
In figure \ref{fig:feyncc11} the 10 diagrams contributing at tree level to the 
e$^+$e$^-\to u\bar{d} \mu\bar{\nu_\mu}$ events are shown. 
All semileptonic decays (except those where an 
electron is present in the final state) are produced through the same set
of diagrams, with the proper redefinition of the final state particles.
The graphs 1, 6 and 7 are the only ones where a W pair is produced; 
they are often referred to as CC03 processes.
Contributions from single- and non-resonant processes are particularly 
large for $l\nu l\nu$ and $qqe\nu$ final states, leading to an ambiguity in
the definition of the signal. The approach followed by the LEP collaborations
is slightly different:\begin{itemize}
\item   DELPHI, L3: consider efficiencies on signal inside generator level 
cuts, and apply multiplicative factors for translating the cross section
measured for the full set of diagrams into a cross section relative to
W-pair production only
\item   ALEPH, OPAL:   consider the additional diagrams as a background, 
neglecting the interference between these processes and the W pair diagrams
\end{itemize}
It was shown that these procedures give the same results within a 1\% accuracy.

\section{Machine parameters, schedule and calibration}
Due to the strong increase in synchrotron radiation, the only possibility to
operate the LEP machine well above the Z resonance is to considerably increase
the LEP1 accelerating power. Since the machine layout cannot be changed,
this can only be achieved raising the accelerating gradient in the straight
sections. The 128 five-cell copper cavities constituting the accelerating system
of LEP in the first phase were able to deliver a peak RF power corresponding
to a voltage of 400 MV
per revolution, clearly inadequate for LEP2 needs (over 3000 MV). The big
increase in performances was only possible due to the operation of 
superconducting cavities, that because of their very high quality factor could
provide as much as 6 MV/m of accelerating gradient.\par
The installation of those cavities proceeded in several steps, and so did
the energy of operation of the machine. The machine schedule in the period
1996-1997 is shown in table \ref{tab:sched}, where only the data-taking periods
with total energy above the Z peak have been considered. Apart from the runs above 
the WW threshold, relevant to this report, a short run at 130-136 GeV has been
taken, to clarify possible anomalies in the 4-jet production at that energy.
The results of that run are summarized in \cite{4jet}.
\begin{table}
\begin{center}
\begin{tabular}{|l|l|l|l|l|}\hline
Period&N. SC cavities&N. Cu cavities&$\int{\cal L}$ (pb$^{-1}$)&
Energy (GeV)\\ \hline
1996 a&144&182&12.1&161\\
1996 b&176&150&11.3&172\\
1997&240&86&63.8&183\\
1997 b&240&86&7.2&130-136\\ 
1998 & 272&48&196.4&189\\ \hline
\end{tabular}
\end{center}
\label{tab:sched}
\caption{Characteristics and performances of the LEP machine in the years 1996-1998. Only runs above the Z peak have been listed.}
\end{table}
\par
Since all fitting methods use the centre-of-mass energy as a kinematic
constraint, the uncertainty on the knowledge of the LEP beam energy directly 
reflects into a systematic error on the value of the W mass:
\[\Delta M_W=\frac{\Delta E_b}{E_b} M_W\] 
To fulfill the
requested precision on the mass, the LEP energy has to be known with an 
accuracy better than 20 MeV. At LEP1, energy calibration was performed via
resonant depolarization (RDP)\cite{respol}. This method can not be used at 
LEP2, since there is no possibility to have polarization at physics energies.
Several RDP measurements have been however performed at lower energies,
and extrapolated. The main error on this method comes from the extrapolation
itself, leading to a total error of about 20 MeV at 189 GeV.\par
As an independent approach, a spectrometer is planned for LEP, to be 
operational in 1999 and beyond. This will consist in a fully equipped dipole
magnet, giving a precision of about $1 \mu m$ on the beam position, extracting
the particle momentum out of their curvature in the dipole magnetic field.
\par
\section{WW cross section and branching ratios}
WW events can have very different topologies, depending on the different
decays of the two W bosons. As two extreme cases, the decay of two W bosons
could produce a high-multiplicity four-jet event, as well as a low-energy
imbalanced event with only two charged leptons seen in the detector.\par
Accordingly, the selection criteria, the backgrounds and the possible
systematic uncertainties in the selections can be very different. All LEP
collaborations have different analysis for the possible WW decay channels.
They are here grouped into three main categories: fully leptonic, semileptonic
and fully hadronic decays.

\subsection{Fully leptonic events}
\begin{figure}[tb]
 \begin{minipage}{.45\linewidth}
  \begin{center}
   \includegraphics[width=6cm,bb=30 220 500 640,clip]{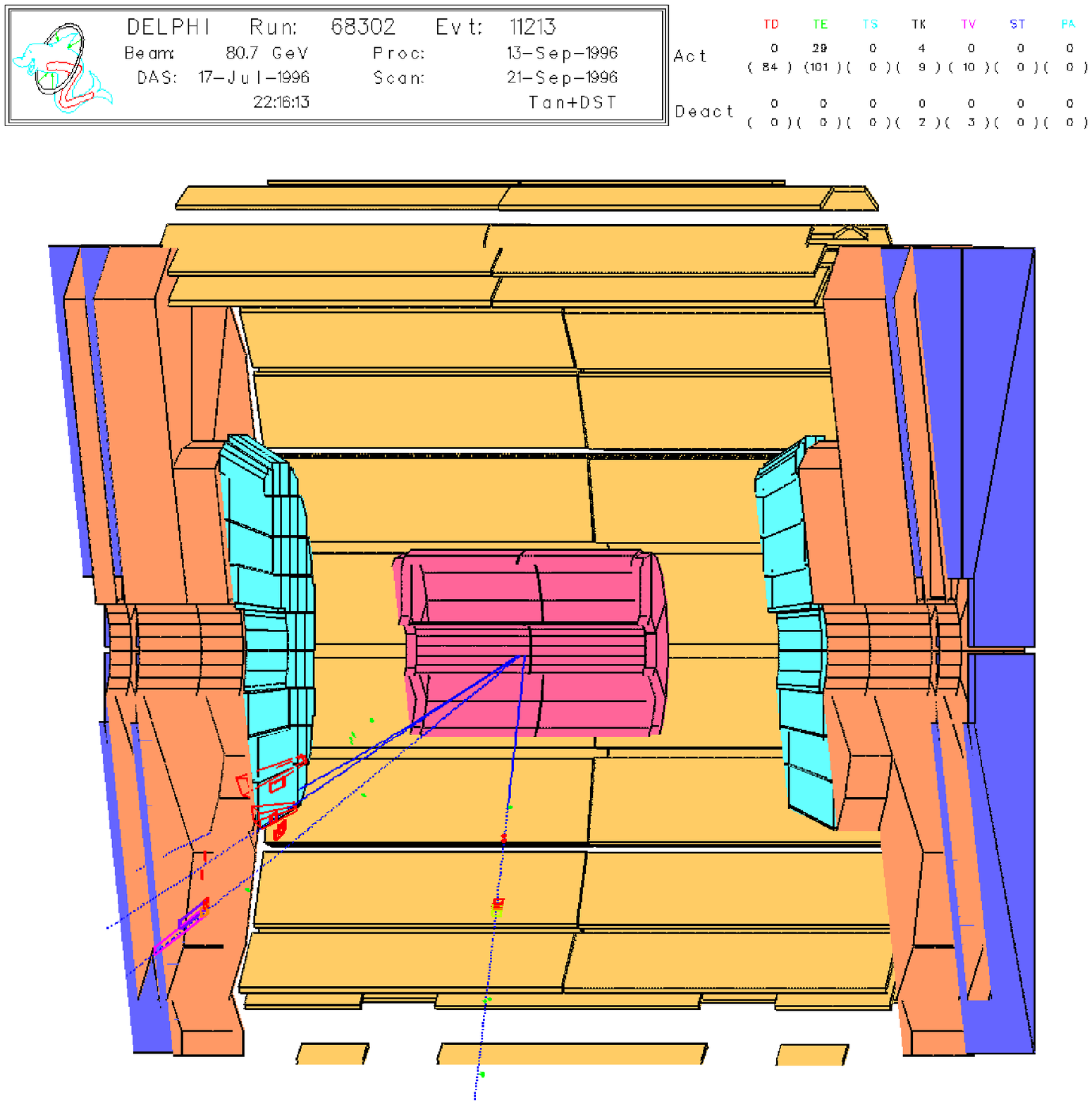}
  \caption{A $WW\to \tau\nu \mu\nu$ event in Delphi. A narrow jet originated
from $\tau$ decay (on the left), and a muon (traversing the whole detector)
are widely acoplanar.} \label{fig:delpevlnln}
  \end{center}
 \end{minipage}
 \begin{minipage}{.45\linewidth}
  \begin{center}
   \includegraphics[width=7cm]{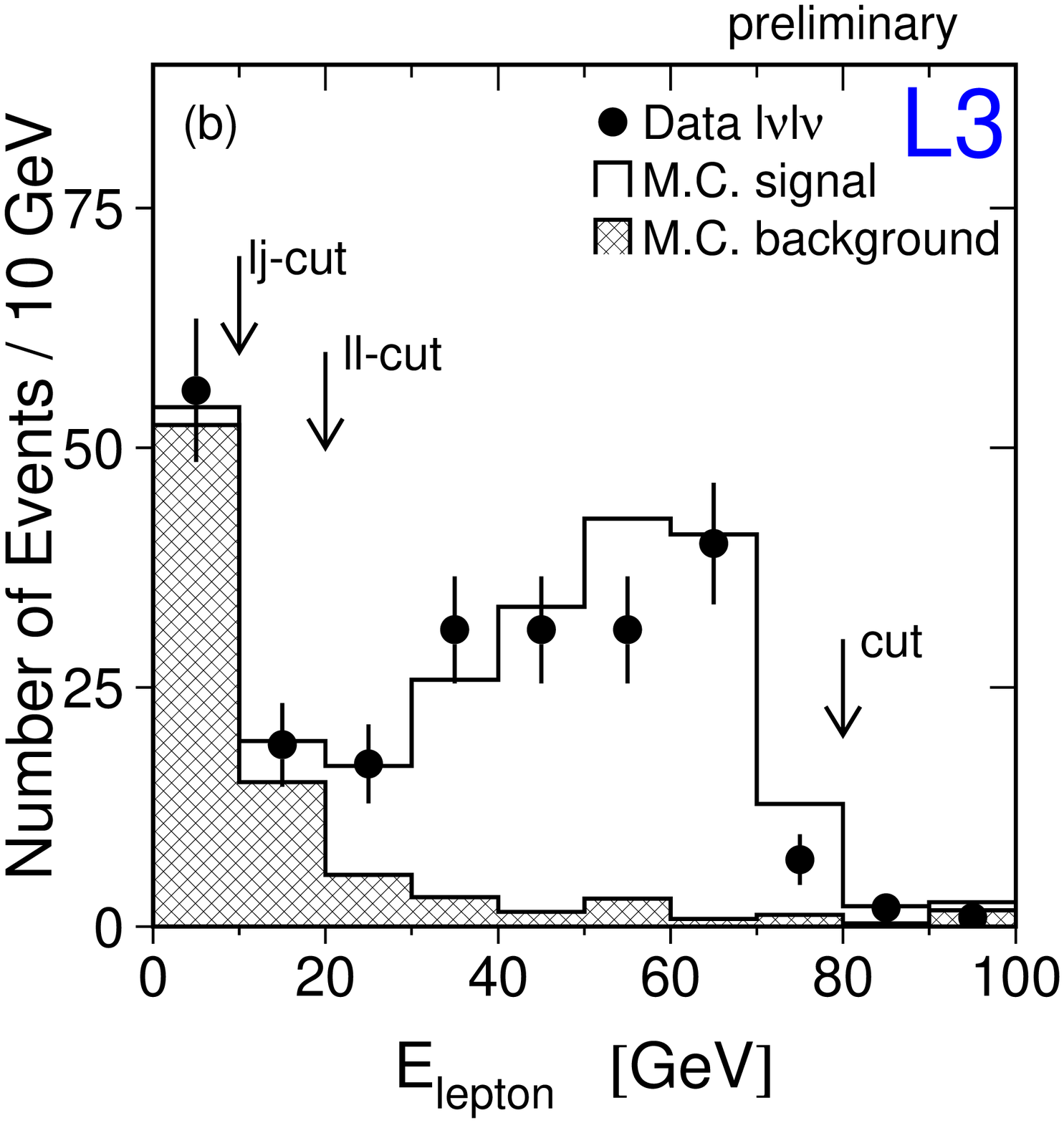}
  \caption{Lepton energy distribution in L3 for fully leptonic events} \label{fig:l3elept}
  \end{center}
 \end{minipage}
\end{figure}

Fully leptonic events $WW\to l\nu l\nu$ are usually characterized by:
\begin{itemize}
\item two high-energy acoplanar leptons
\item missing momentum not pointing to the beam pipe, due to the undetected
neutrinos
\end{itemize}
An example of a $l\nu l\nu$ event detected in the DELPHI detector is shown
in figure \ref{fig:delpevlnln}. The energy distribution of the most energetic
lepton in L3 for $l\nu l\nu$ candidates is  in figure \ref{fig:l3elept}.\par
Events are usually classified into lepton-lepton, lepton-jet and jet-jet
categories, where here lepton stands for either electron or muon, and narrow
jets are considered, to account for hadronic $\tau$ decays.\par
\par
Due their topology, the main backgrounds to these processes are:
\begin{itemize}
\item Two-fermion events from $Z/\gamma$ decays (especially $\tau$ pair
production), Bhabha scattering events
\item high-energy $\gamma-\gamma$ interactions
\item $ZZ\to ll\nu\nu$ events (mainly for $\sqrt{s} > 184$ GeV)
\end{itemize}
\par
Since the leptons are required to be acoplanar, the most dangerous background
from two fermion events is represented by radiative $Z/\gamma$ decays.
Events with a high-energy isolated photon are rejected, since this is a
clear indication of radiative $Z/\gamma$ decays. Also events with missing
momentum pointing at very low angle are rejected, since in this case the
radiated photon could have been lost in the beam pipe or in a badly-instrumented
sector of the detector.
The typical selection efficiencies for this channel are higher for the case
in which two stable leptons are produced, and lower for the jet-jet case, due
to the stronger cuts needed to suppress the larger background.
Overall efficiencies are around 70\%. 
\par
\subsection{Semileptonic events}
The semileptonic channels $WW\to q\bar{q}l\bar{\nu}_l$, in particular those 
with an electron or a muon in the final
state, have quite similar topology. These events are characterized by two
hadronic jets, a high-energy lepton and large (and similar) missing energy 
and momentum due to the neutrino. $qq\tau\nu$ events are usually more balanced
due to the additional neutrinos produced in $\tau$ decays, and the missing
energy is larger. The lepton from $\tau\to e$ and $\tau\to\mu$ decays is 
softer, than that produced directly from the W, while hadronic $\tau$ decays
produce a narrow jet.\par
The background is mainly coming from hadronic Z radiative decays, but is
different for the three channels. For the $qqe\nu$ case, the main baground 
comes from Z radiative events with the photon in the detector, associated to a
nearby track, or converted into a $e^+ e^-$ pair. This is particularly
true in the forward region of the detector, where most of the radiative photons
are emitted, and where usually the tracking capabilities of the detector are 
not optimal. Background to the $qq\mu\nu$ channel is mainly coming from 
semileptonic decays of b quarks in $Z\to qq (\gamma)$ events, or from $ZZ\to
qq\mu\mu$ processes, where one of the muons is not identified, and mimics
missing momentum. Having less strong signature, the $qq\tau\nu$ channel has
usually more complicated selections, and its main background arises from 
radiative Z hadronic decays where the photon gets undetected and a third
jet fakes that coming from $\tau$ decays.\par
For the final selection, DELPHI and L3 use a cut-based approach, requiring
good isolation for the lepton and high invariant mass for the decay products
of the two Ws. ALEPH and OPAL combine the informations coming from similar
variables using an event probability function. Typical efficiencies are
of the order of 80\% for the electron and muon channels, and 50\% for the
$\tau$ channel.\par
\begin{figure}[tb]
  \begin{center}
   \mbox{\includegraphics[width=6cm,bb=100 228 534 652,clip]{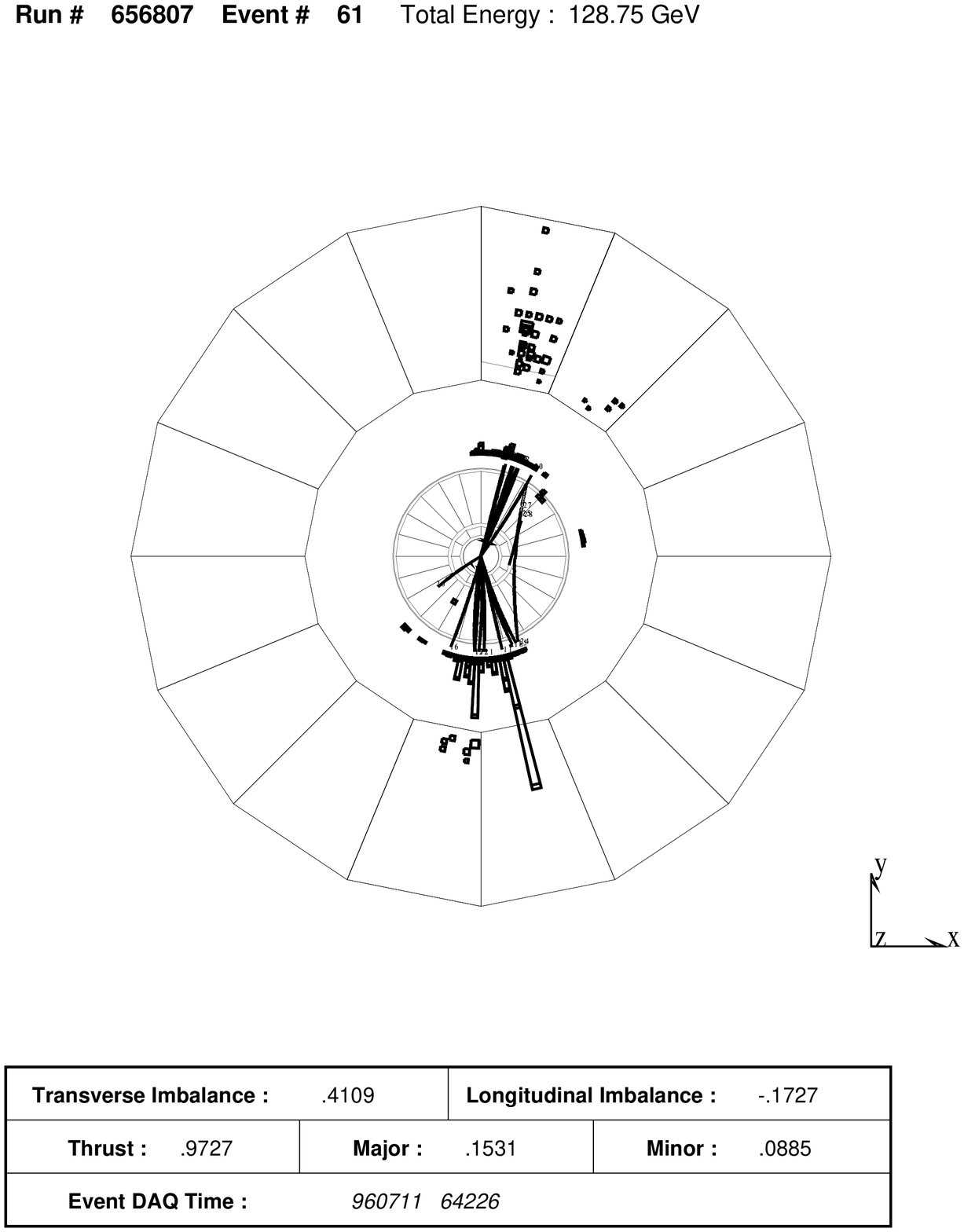}}
  \end{center}
  \caption{A $qqe\nu$ event in L3. The electron is visible as a large ``tower''
in the electromagnetic calorimeter, in the bottom part of the event. The
two hadronic jets are opposite to each other.}
\label{fig:qqlnfig}
 \end{figure}
 \begin{figure}
  \begin{center}
   \includegraphics[width=14cm]{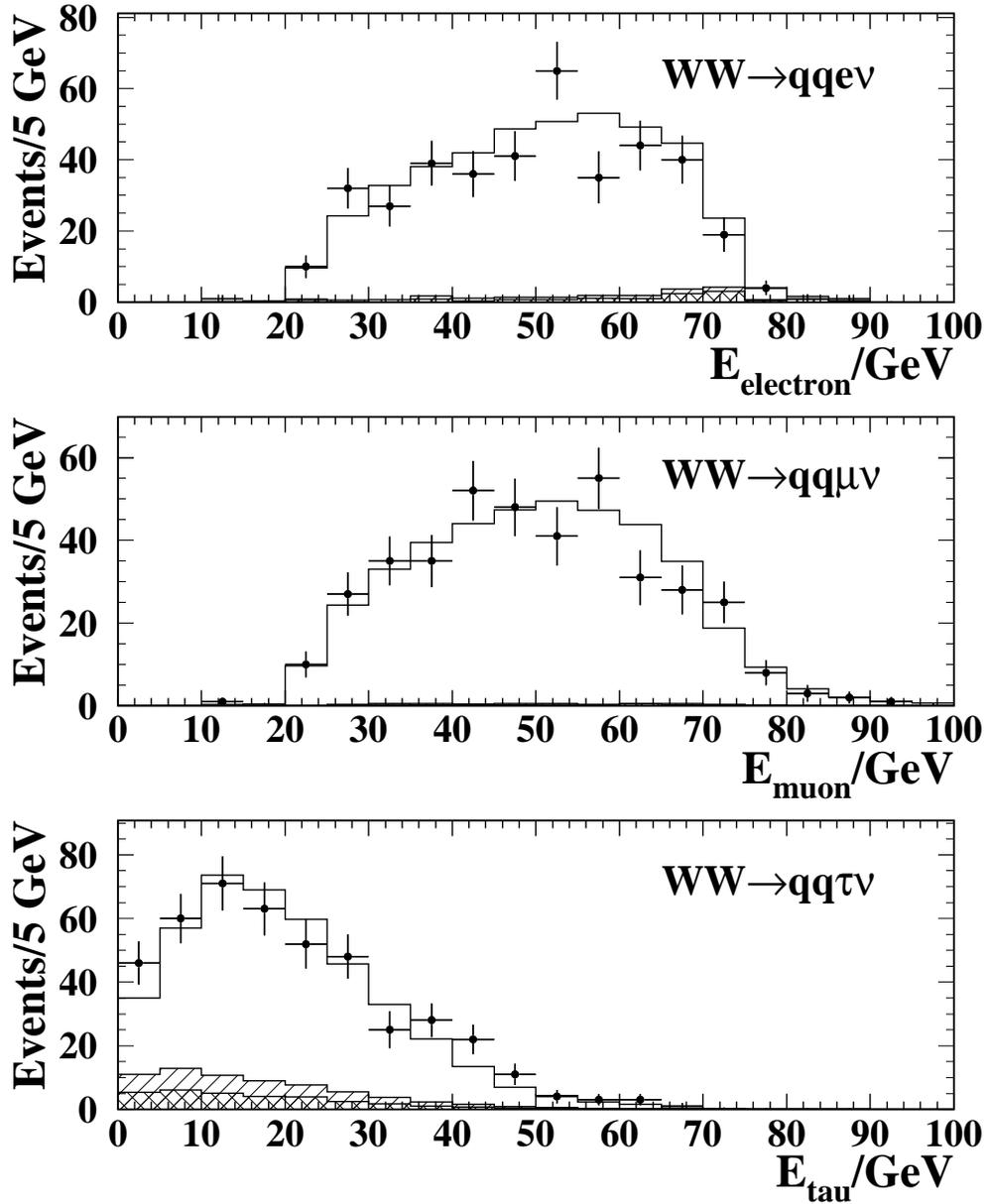}
  \end{center}
  \caption{Distributions of measured lepton energies for events selected
in OPAL in the three semileptonic channels} \label{fig:qqlndist}
\end{figure}
\par
\subsection{Hadronic W decays}
The channel $WW\to q\bar{q} q\bar{q}$ has about the same branching ratio as
the sum of three semileptonic ones, i.e. about half of the total number of
WW decays. It is characterized by four high-energy well separated hadronic 
jets, coming from hadronization of the quarks from the W decays.\par
There are two main sources of background:\begin{itemize}
\item $Z\to q\bar{q}$ events with hard gluon radiation
\item $ZZ\to q\bar{q} q\bar{q}$ events.
\end{itemize}
The latter is almost irreducible, since well-isolated high energy jets
are produced, and the only difference with respect to the signal is the
slightly higher jet-jet invariant mass. Z decays have a larger cross section,
but since the two additional jets are coming from gluon radiation, they are
usually less energetic and closer to the emitting quark.\par
\begin{figure}[p]
  \begin{center}
   \includegraphics[width=14cm]{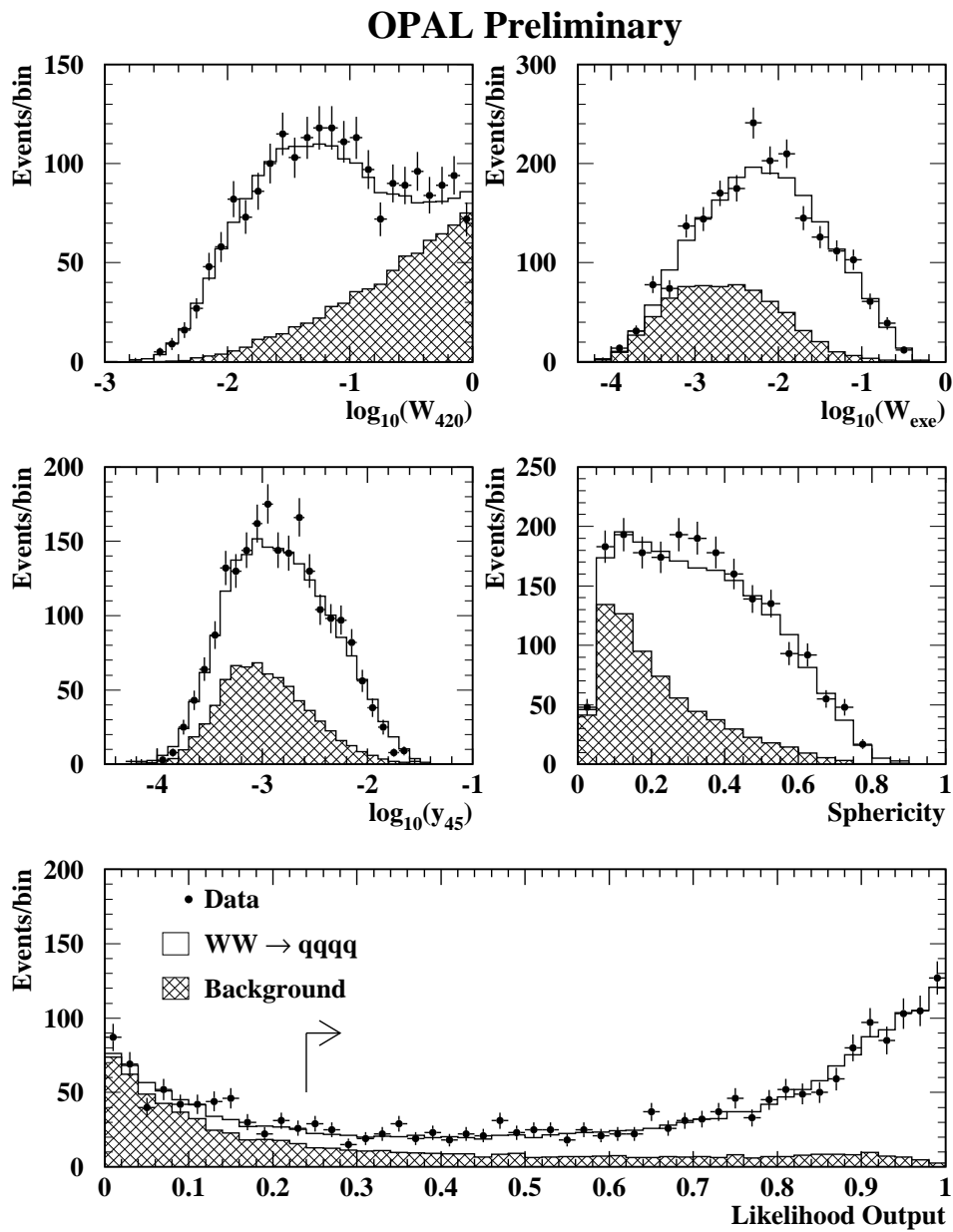}
  \end{center}
  \caption{Distributions of the variables used by OPAL in the $WW\to q\bar{q}
q\bar{q}$ analysis. The points indicate the data, the open histogram
represents the MonteCarlo expectations for the signal, and the hatched
histogram shows the background estimate.} \label{fig:opalqqqq}
\end{figure}

All experiment try to combine all available informations in an optimized way.
This is done combining several variables (event shape, invariant 
masses angles between jets etc.), using a likelihood discriminator (OPAL) or 
a neural network (DELPHI,ALEPH, L3). 
The variables used by the OPAL collaboration are shown in figure 
\ref{fig:opalqqqq}, together with the resulting likelihood and the value of the
cut. In order to have an additional gain in statistical power and have a
further cross-check on the background, ALEPH and L3 fit the neural network 
output distribution by a linear combination of distributions for signal and 
background. 
\par
\begin{figure}[tb]
  \begin{center}
   \includegraphics[width=8cm,angle=270,bb=40 30 570 800,clip]{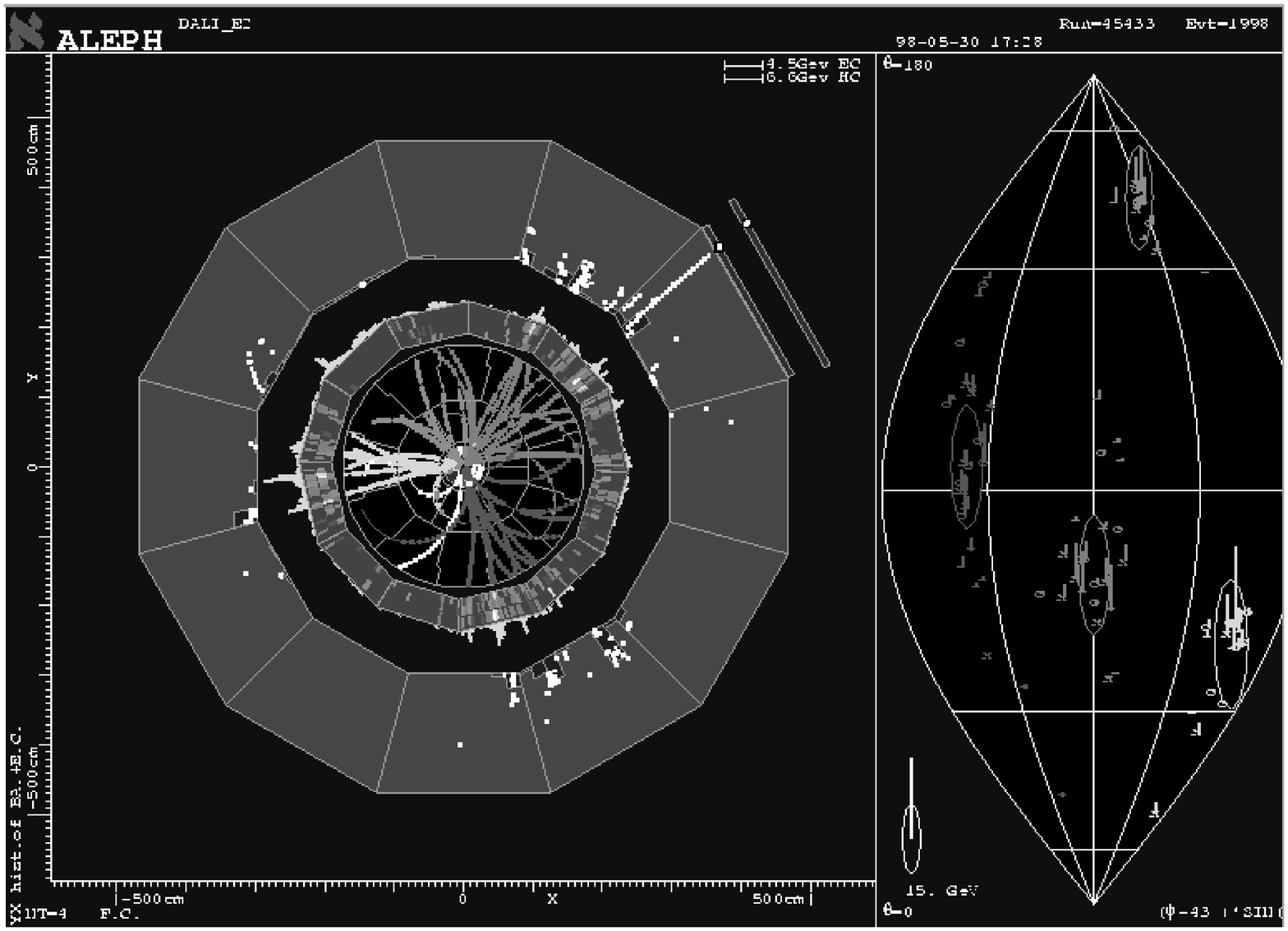}
  \end{center}
  \caption{A $WW\to q\bar{q} q\bar{q}$ event in ALEPH. The four jets coming
from W decays are clearly separated.}
\label{fig:al4j}
 \end{figure}
The uncertainties related to the QCD modeling of the fragmentation process,
in particular of the final state interactions, are the main sources of
systematic errors for this channel. These uncertainties will be discussed in 
more detail in the section about W mass measurements.
\par
\section{Determination of WW cross section and branching fractions}
The cross sections for the individual WW decay channels measured as above
are combined to extract a value of the total WW production cross section
and branching ratio. To make use of physical assumption like i.e. lepton 
universality, likelihood-based fit are used by the different collaborations.
For a given channel $i$, the expected number of events, $\mu_i$, is computed
accounting for the background and the cross-efficiencies among that channel
and all the others ($\epsilon_{ij}$):
\[\mu_i=L\times(\Sigma_j \epsilon_{ij}\sigma_j+\sigma_i^{bg})\]
where L is the collected luminosity and the sum runs on all channels.
The cross sections are extracted maximizing the likelihood
\[{\cal L}=\Pi_i P(N_i,\mu_i)\]
where P is the Poisson probability of observing $N_i$ events in a given 
channel i, with $\mu_i$ expected.\par
This likelihood can be maximized leaving different free parameters, according
to the physics assumptions. If for instance no assumptions are made, the 
cross sections $\sigma_j$ for all channels are left free. On the other hand,
it is possible to extract the total cross section, imposing the knowledge of
the W standard model branching fractions. In this case, the above cross 
sections are expressed as
the product of the WW cross section $\sigma$ (which is now the only free 
parameter in the fit) times the branching ratio for the
corresponding channel.\par
The total WW production cross section for the four experiments at $\sqrt{s}$
of 183 and 189 GeV are listed in table \ref{tab:totxsec}. All results from the
run at 189 GeV are preliminary, and taken from the contributions of the
various collaborations to the winter conferences\cite{xsecmor99}.\par
\vspace{1cm}
\begin{table}
\begin{center}
\begin{tabular}{|l|l|l|}
\hline
\multicolumn{3}{|c|}{Cross-section $\sigma_{\tt CC03}$ (pb)}\\
\hline
Experiment  & $\sqrt{s}=$183~GeV & $\sqrt{s}=$189~GeV\\
\hline
ALEPH  & 15.57 $\pm$ 0.68$^n$ & 15.64 $\pm$ 0.43 \\
DELPHI & 15.86 $\pm$ 0.74$^n$ & 15.79 $\pm$ 0.49 \\
L3     & 16.53 $\pm$ 0.72$^p$ & 16.20 $\pm$ 0.46 \\
OPAL   & 15.43 $\pm$ 0.66$^p$ & 16.55 $\pm$ 0.40 \\
\hline
LEP &  15.83 $\pm$ 0.36 & 16.07 $ \pm$ 0.23 \\
 SM &  15.70 $\pm$ 0.31 & 16.65 $ \pm$ 0.33 \\
\hline
  \multicolumn{3}{|c|}{$^p$ Published $^n$ New  Preliminary}\\
\hline
\end{tabular}
\end{center}
\label{tab:totxsec}
\end{table}
\par
\vspace{1cm}
\par
\begin{figure}[p]
  \begin{center}
   \includegraphics[width=6cm]{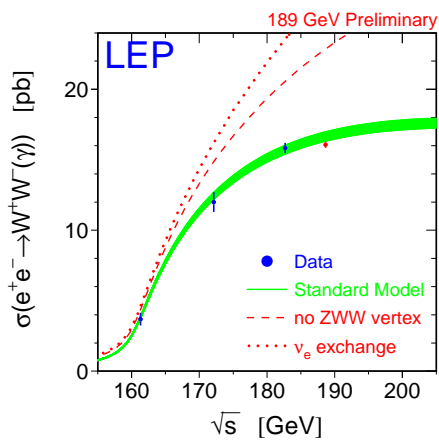}
  \end{center}
  \caption{Total WW production cross section} \label{fig:xsec}
\end{figure}

The combined LEP cross section only considers statistical errors. The combined
value for all LEP2 energies is shown in figure \ref{fig:xsec}. In addition to
the curve predicted from the standard model, this figure shows the WW cross
section in the two cases where the ZWW vertex has zero coupling, and where
WW production occurs only via the neutrino exchange diagram. In both cases
the WW production cross section diverges for large values of $\sqrt{s}$,
and is also incompatible with the values measured at LEP. Therefore, the
simple cross section measurement represents a confirmation of the non-Abelian
structure of the electroweak interactions.
\begin{figure}[p]
 \begin{minipage}{.45\linewidth}
  \begin{center}
   \includegraphics[width=6cm]{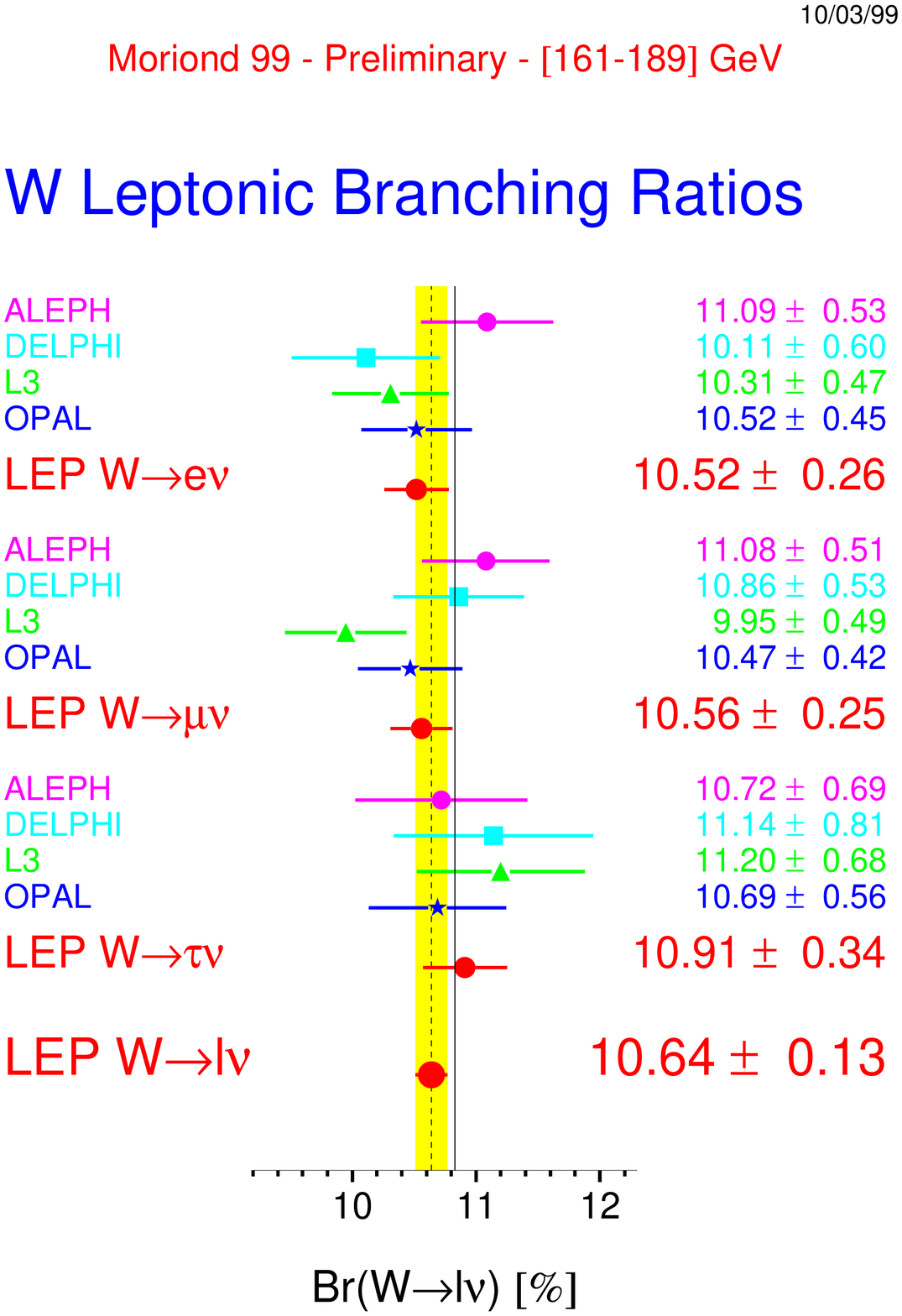}
  \caption{Leptonic branching ratios} \label{fig:lbr}
  \end{center}
 \end{minipage}
 \begin{minipage}{.45\linewidth}
  \begin{center}
   \includegraphics[width=6cm]{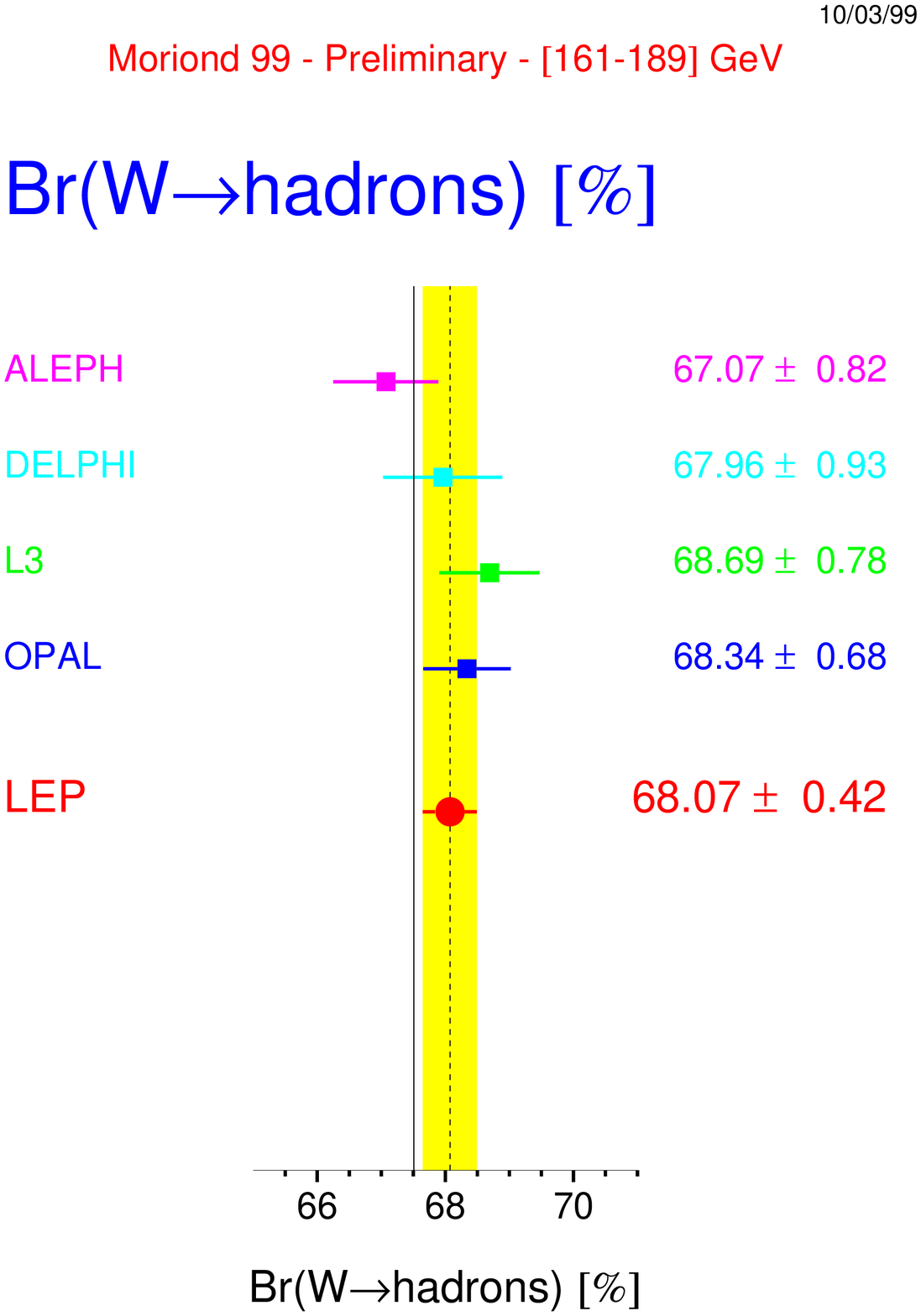}
  \caption{Hadronic branching ratios} \label{fig:hbr}
  \end{center}
 \end{minipage}
\end{figure}

If the cross section is not fixed, it is possible to
determine the W decay branching ratios, leaving them as free parameters for 
the fit. The results for the different leptonic ratios is shown in figure
\ref{fig:lbr}, showing a direct verification of th lepton universality
in charged current weak interactions. The hadronic branching ratio is shown
in figure \ref{fig:hbr}. This value can be expressed in terms of the CKM 
matrix elements using the following expression:
\[\frac{Br(W\to q\bar{q})}{1-Br(W\to q\bar{q})}=(1+\frac{\alpha}{\pi})\Sigma
|V_{ij}|^2\]
that yields, using the combined LEP value:
\[\Sigma |V_{ij}|^2=2.10\pm 0.08\]
The experimental knowledge of all elements of the CKM matrix is quite good,
apart from $V_{cs}$, suffering from large uncertainties (about 20\%) of both 
experimental and theoretical nature\cite{pdg}. 
Imposing unitarity of the CKM matrix
and considering the measurements of the other matrix elements, the previous
result can be reinterpreted as a determination of $|V_{cs}|$: 
\[|V_{cs}|=1.002\pm 0.0016 (stat) \pm 0.002 (syst)\]  
A completely independent technique to determine this quantity will be presented
in section 12.\par
\par
\section{W mass}
As mentioned in the introduction, the W mass is one of the most important
measurements of the LEP2 program. At energies close to the WW production
threshold, the highest sensitivity is reached deriving the mass from the
cross section measurement, an approach conceptually similar to that used
at LEP1 to measure the mass of the Z boson. At higher energies, the
W mass is derived directly from the measured invariant mass of the W decay
products.\par
\subsection{W mass from threshold cross section}\par
\begin{figure}[tbh]
  \begin{center}
   \includegraphics[width=8cm,clip]{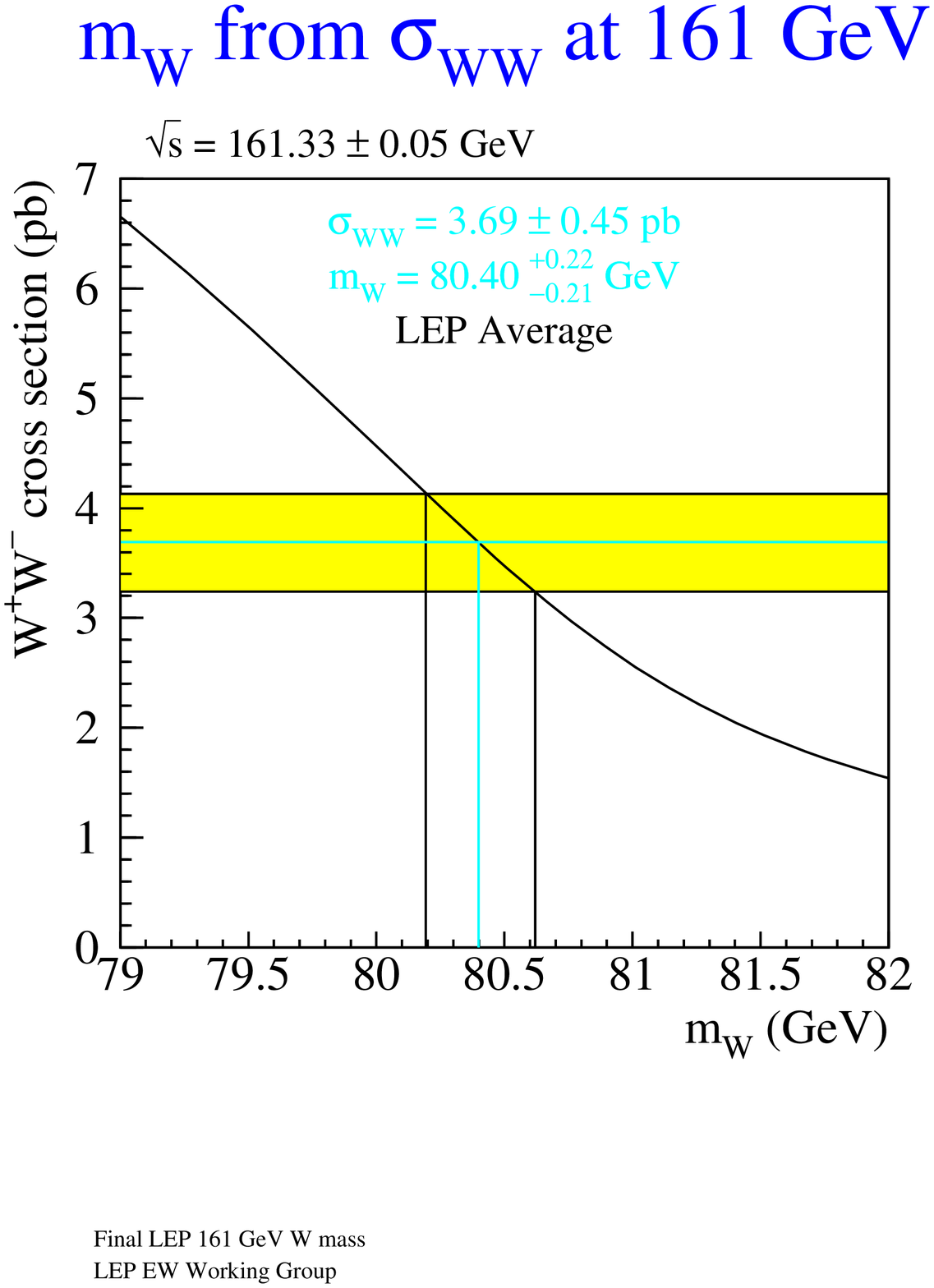}
  \end{center}
\label{fig:mass161}
\caption{W mass from the combined LEP cross section at $\sqrt{s}=161$ GeV}
\end{figure}\par
Assuming validity of the SM, the W mass can be extracted from the cross 
section, for a fixed centre-of-mass energy. The sensitivity of this approach 
is maximal close to threshold, due to the steep rising of the cross section
in that region; for this reason this method was used to determine the mass
from the cross section of the first run at $\sqrt{s}=161$ GeV. In figure
\ref{fig:mass161} the dependence of the cross section on the W mass is 
shown, together with the combined measurement of the mass from the LEP
experiments.\par
\subsection{W mass from direct reconstruction}
At higher energies the dependence of the cross section on the W mass is
negligible, so the derivation of the mass from the cross section can no 
longer be used. On the other hand, the
WW cross section is much larger than at threshold, and it is possible to use
the direct reconstruction method, i.e. the W mass is extracted with a fit to 
the invariant mass distribution of the W decay products. The calculation of 
these masses is only trivial in the semileptonic case, where two jets are 
coming from a W and the system of lepton and neutrino from the other.
In the fully leptonic case, the system is underconstrained, due to the
presence of at least two undetected neutrinos. The ALEPH collaboration has
been the only one so far to use this channel for mass fits, using the energy
of the two leptons, with small statistical power. In the fully hadronic
case, since at least four jets are present in the detector, several mass pairs
could be formed. Criteria based on reconstructed masses, angles etc.
are used to get the best pairing, with efficiencies of the order of 80\%;
however including the other pairings with smaller weight can help increasing
the mass sensitivity.\par
The main problem to measure the W mass from reconstructed distribution is
to account for all distortions coming from detector effects and selection
biases. Two main approaches are used:\begin{itemize}  
\item convolution
\item MC reweighting
\end{itemize} 
All LEP collaborations use both methods, quoting one for the final results 
and the other as a cross-check.\par
In both cases the final error on the W mass will be determined by the total
number of candidates as well as the detector resolution in measuring invariant
masses. In order to improve the detector performances, it is possible to 
impose some physical constraint to any single event. Since energy and momentum
are conserved in the collision, it is possible to perform a fit to energies
and angles of the final states particles, such that the fit results satisfy
the kinematical constraint and are as close as possible to
the measured ones. This procedure largely increases the sensitivity to the W
mass, but requires a precise knowledge of the LEP energy since this value is
directly used in imposing the energy conservation.\par 
The L3 and OPAL collaboration also exploit the additional kinematical 
constraint that the masses of the two produced W bosons must be
equal within the W width.\par
 
\subsection{Convolution}
\begin{figure}[tbh]
  \begin{center}
   \includegraphics[width=9cm]{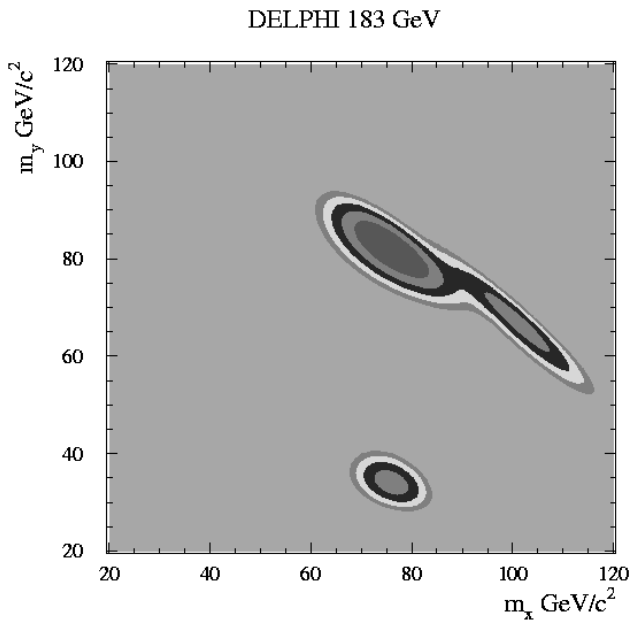}
  \end{center}
  \caption{Likelihood contours in the $M^1_W-M^2_W$ plane for a four-jet 
event in DELPHI. The three maxima correspond to the three possible
jet pairings.}
\label{fig:delpev}
\end{figure}

In the convolution method (DELPHI), the theoretical W line-shape curve 
(depending on the W mass) is convoluted with an analytical function 
describing detector effects. The experimental line-shape is compared to the
convoluted curve, determining a likelihood function, having as a free
parameter the W mass. Maximizing the likelihood it is possible to extract
the value of the W mass for which the curve obtained smearing the theoretical
distribution mostly resembles the experimental curve.\par
The main difficulty of this method is in the modelization of the the detector 
response, that must be included an analytical form. Morover, the reconstructed
mass has a bias that must be corrected comparing with the MC. On the other 
hand, the method allows the use of different event weights depending on the 
detector resolution for each data event, thus improving the accuracy of the
measurement.\par
In particular, DELPHI fits the masses in the $M_W^1$, $M_W^2$ plane, using
all three combinations for the qqqq channel (see figure \ref{fig:delpev}).
\par
\subsection{MC Reweighting}
\begin{figure}[p]
 \begin{minipage}{.45\linewidth}
  \begin{center}
   \includegraphics[width=6cm]{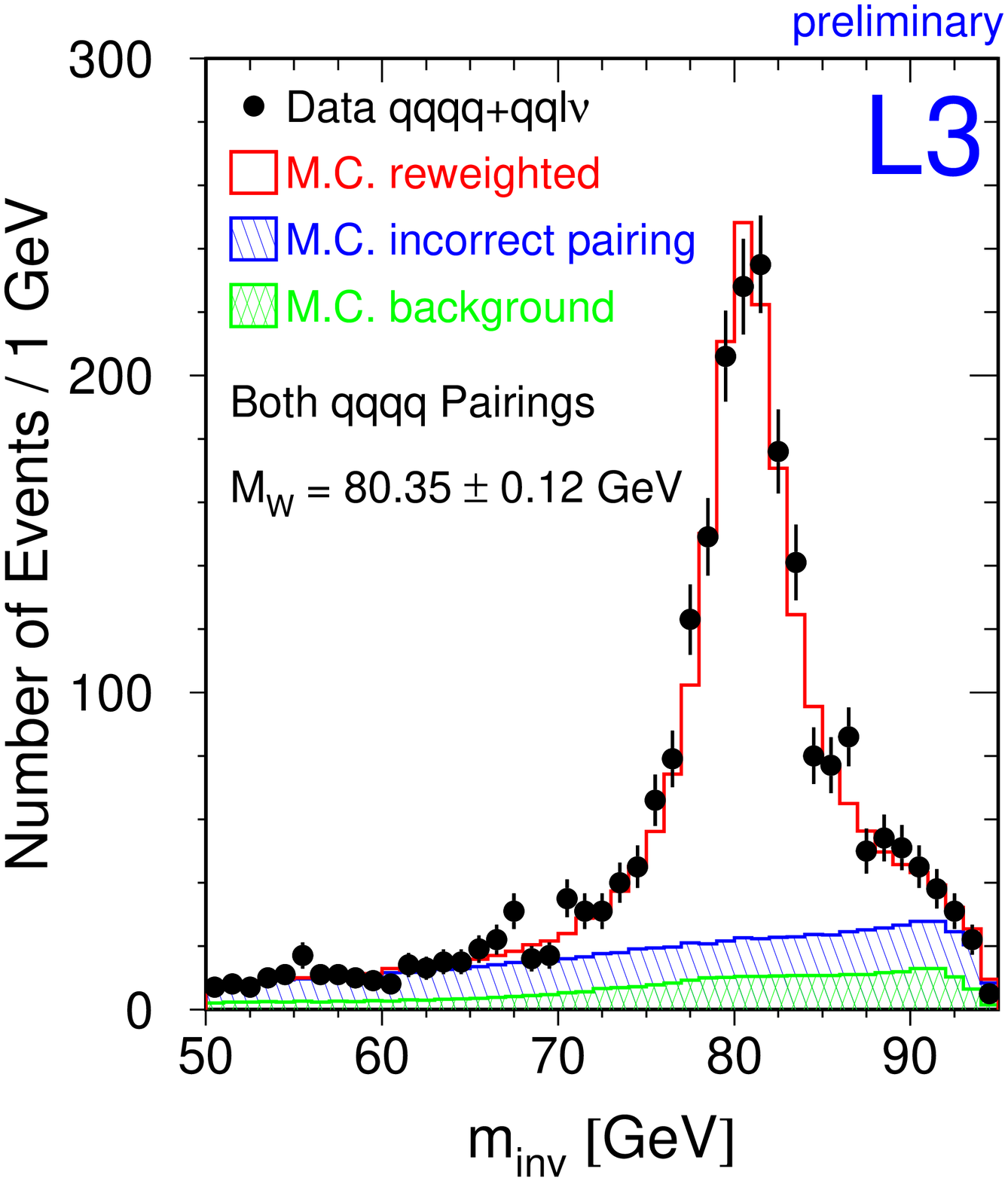}
  \caption{W mass distribution in L3} \label{fig:l3wmass}
  \end{center}
 \end{minipage}
 \begin{minipage}{.45\linewidth}
  \begin{center}
   \includegraphics[width=6cm]{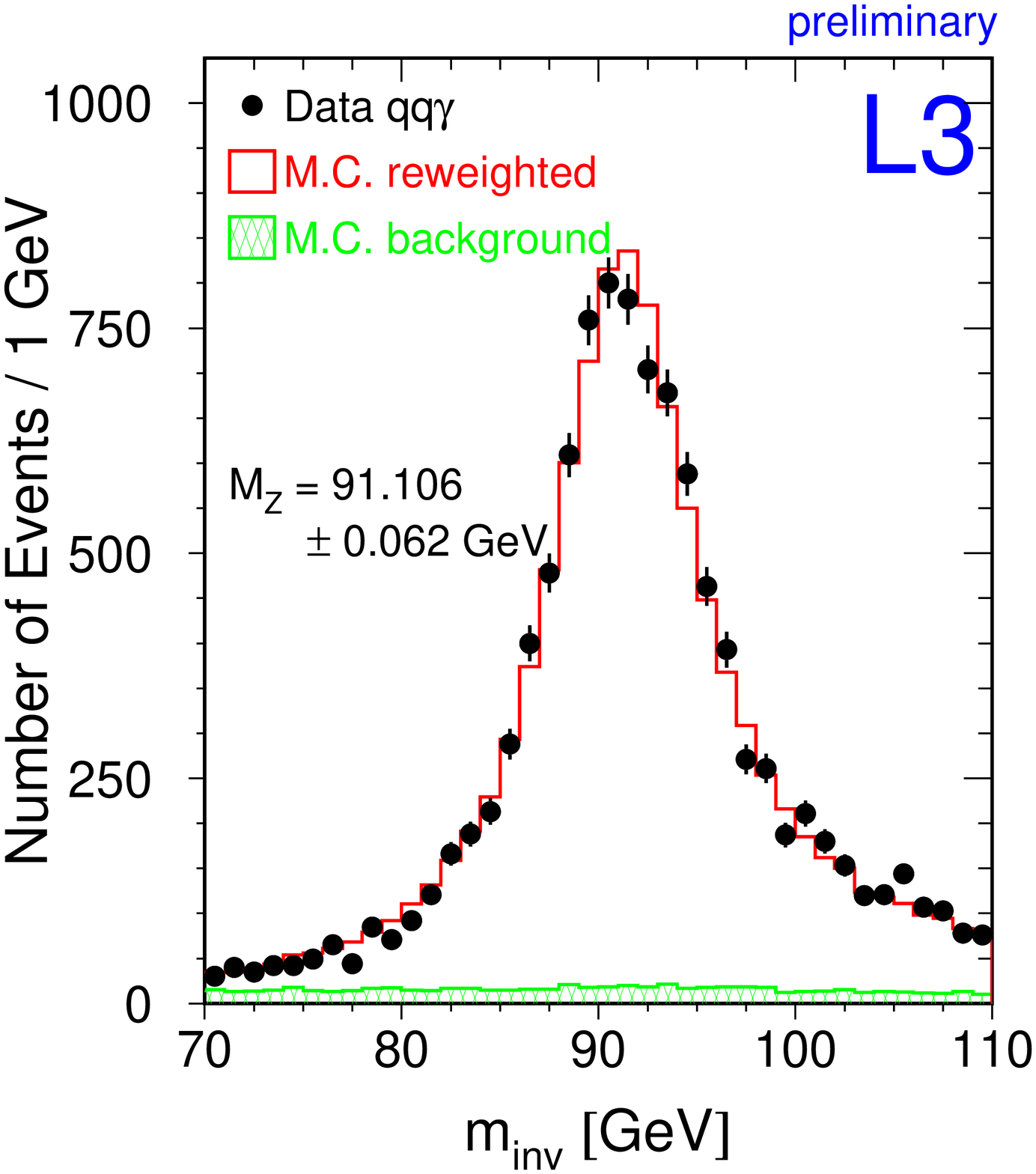}
  \caption{Z mass distribution in L3} \label{fig:l3zmass}
  \end{center}
 \end{minipage}
\end{figure}
\begin{figure}[p]
  \begin{center}
   \includegraphics[width=6cm]{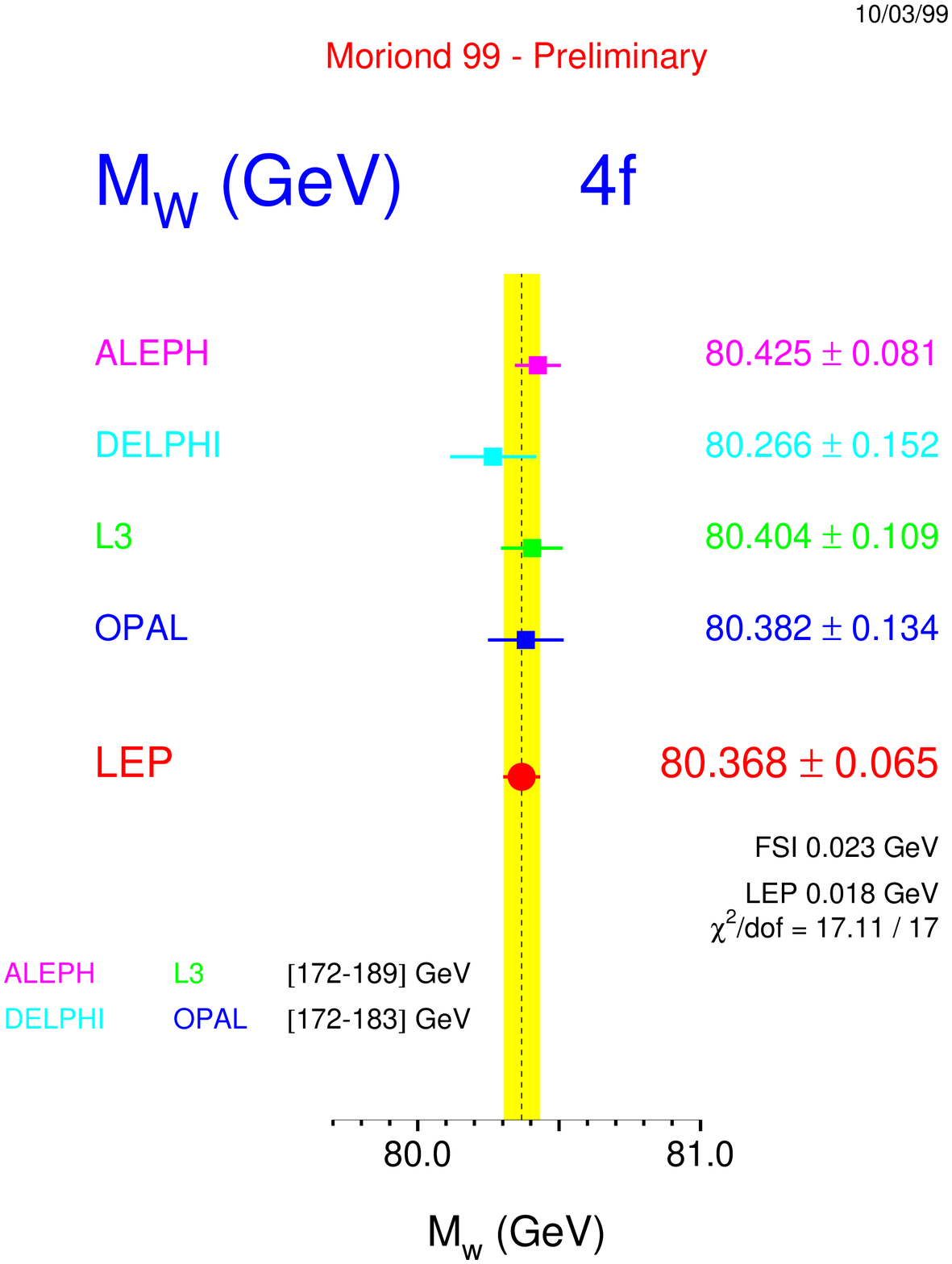}
   \caption{Results for the W mass from direct reconstruction in the four
LEP experiments. ALEPH-L3: preliminary results including the run at 189 GeV;
OPAL-DELPHI: final results from the runs at 172 and 183 GeV of centre of mass
energy}
  \label{fig:wmassres}
  \end{center}
\end{figure}
\par
This method, used by the other collaborations, uses a large number of Monte 
Carlo events to establish the correspondence between generated and 
reconstructed masses. 
These events are generated with a given value of the W mass;
an analytical code is used to reweight them according to the mass that best 
fits the data, using an iterative procedure based on a likelihood similar to 
that used for the convolution method.\par
As a cross-check for the validity of the method, the L3 collaboration has
presented a fit to the Z mass, performed on radiative $Z\to q\bar{q}\gamma$
events, using the same method as the one used to fit the mass of the W. 
Since the value of the Z mass is known with very high precision from LEP1
measurements, the good agreement of the fitted value with the expectation 
is a good test of the complete mass analysis method. In figures 
\ref{fig:l3wmass} and \ref{fig:l3zmass}, the reconstructed invariant mass 
distributions for WW (all channels) and radiative Z events are shown. It is
interesting to notice that for hadronic WW events the two best jet pairing
are included, so a large but flat background from incorrect jet pairing is
present.\par
Both method can be used to perform a two-parameter fit, where both $M_W$ and
$\Gamma_W$ are left free. The correlation of the two measurements is quite
small, and a statistical error of about 200 MeV per experiment on the width
measurement can be obtained.\par
The W mass results from direct reconstruction are shown in figure 
\ref{fig:wmassres}. In this plot, all results refere to data taken at center 
of mass energies between 172 and 189 GeV\cite{massval}.\par
The present combined value from LEP is $M_W=80.368\pm0.065$ GeV (statistical
error only), as precise as the combined value obtained from hadron machines.
\par
\subsection{Systematic uncertainties on the mass measurement}
Given the importance of the measurement of the W mass, as discussed in 
the introduction, and the good statistical accuracy reached by the measurement,
the understanding of the systematic
uncertainties associated to this measurement are crucial to fully exploit the
potentiality of LEP. Systematic uncertainties can come
from several sources:\begin{itemize}
\item beam energy\par
this value is used as a global normalization factor in the kinematic fit,
so its uncertainty directly reflects into an uncertainty on the mass
\item ISR-FSR\par
the incomplete simulations of initial and final state radiation can be 
estimated comparing mass results obtained using different MonteCarlo 
implementations of these effects
\item detector effects\par
errors due to a non-perfect simulation of the detector response can be 
estimated varying resolution and energy scale in reasonable ranges
\item technical effects\par
the finite precision at which the accuracy of the mass fitting method is
tested, as well as the limited MonteCarlo statistics;
\item background\par
the cross section and energy shape of the background is varied, leading to
some small modifications of the measured values of the W mass
\item QCD final state interactions\par
a significant bias to the W mass measured at LEP in the 4-jet channel could
come from QCD interactions in the final state such as color reconnection or
Bose-Einstein effects. Theoretical models for both effects give quite different
results, so presently the experiments assign large systematic errors, comparing
the mass results obtained with the different methods. A more detailed 
description of these effects and of some experimental ways to discriminate
among the various models will be discussed in section 11.
\end{itemize}
In table \ref{tab:systmass}, typical values of uncertainties for the sources
of systematic errors on the W mass listed above are quoted. These numbers
have to be considered as indicative, since they can vary even substantially
from an experiment to another.\par
\begin{table}[h]
\begin{center}
\begin{tabular}{|c|c|}\hline
Effect&Systematic error (MeV)\\ \hline
Beam Energy&20\\
ISR-FSR&10\\
Detector&10\\
Technical&20\\
Background&10\\
Fragmentation&30\\
Final State&30\\ \hline
Total&55\\ \hline
\end{tabular}
\vspace{1cm}
\caption{\label{tab:systmass}Typical valus of uncertainties for the
various systematic sources.}
\end{center}
\end{table}

Presently, the LEP collaborations quote systematic uncertainties larger than
50 MeV (even higher in the 4-jet channel), similar to the present combined
statistical accuracy. Much work is in progress to lower the systematic
unctertainties, in order to fuly profit from the increase in statistics
expected in the next years.\par

\section{Trilinear Gauge Couplings}
\begin{figure}[h]
  \begin{center}
   \includegraphics[width=7cm,bbllx=10,bblly=410,bburx=550,bbury=640,clip]{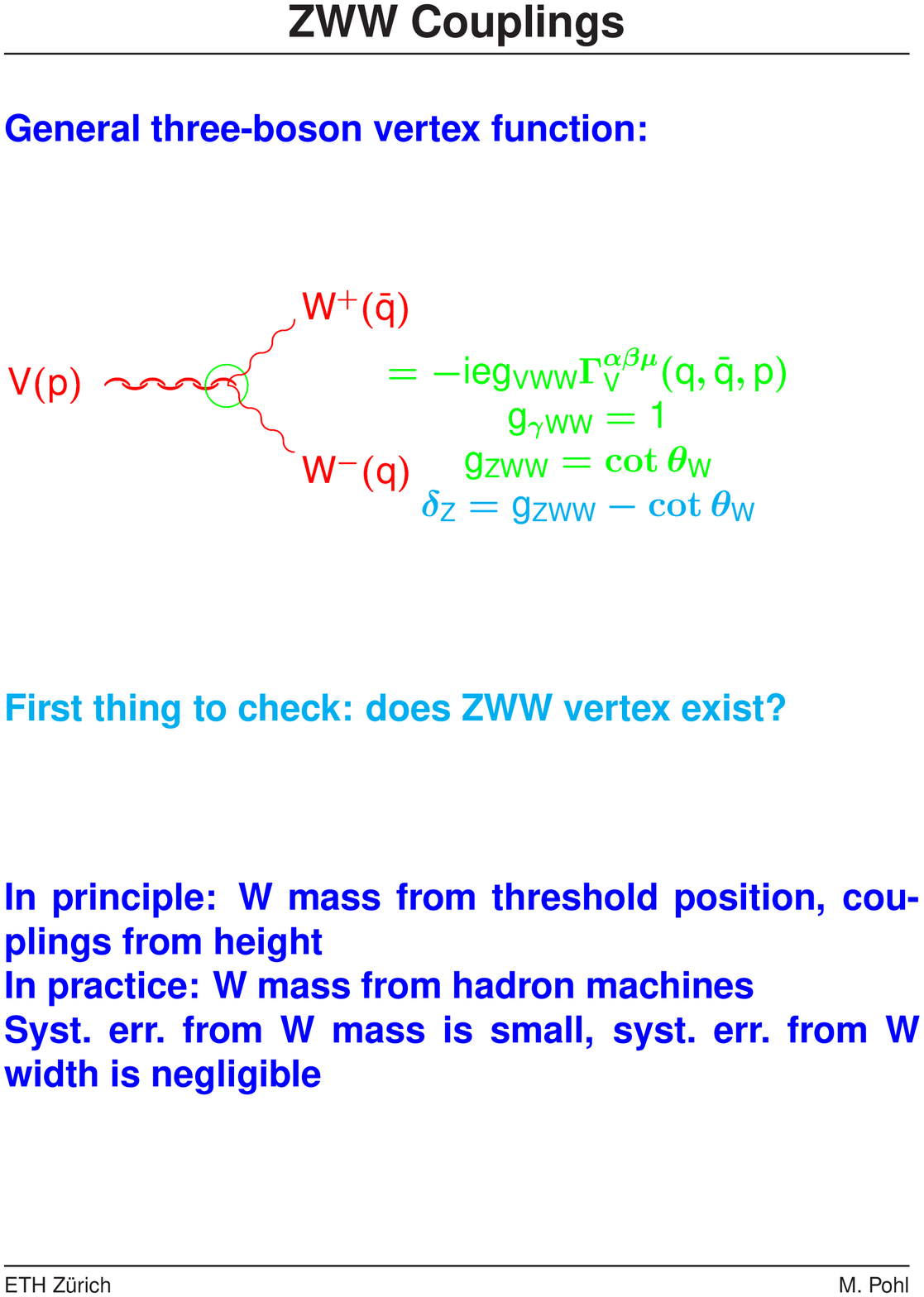}
  \end{center}
\label{fig:tgc}
\caption{Three gauge boson vertex}
\end{figure}
The effect of anomalous trilinear gauge couplings is a modification of the 
WW production cross section and a distortion of the event kinematics.\par
\begin{figure}[h]
  \begin{center}
   \includegraphics[width=7cm,bbllx=30,bblly=190,bburx=570,bbury=600,clip]{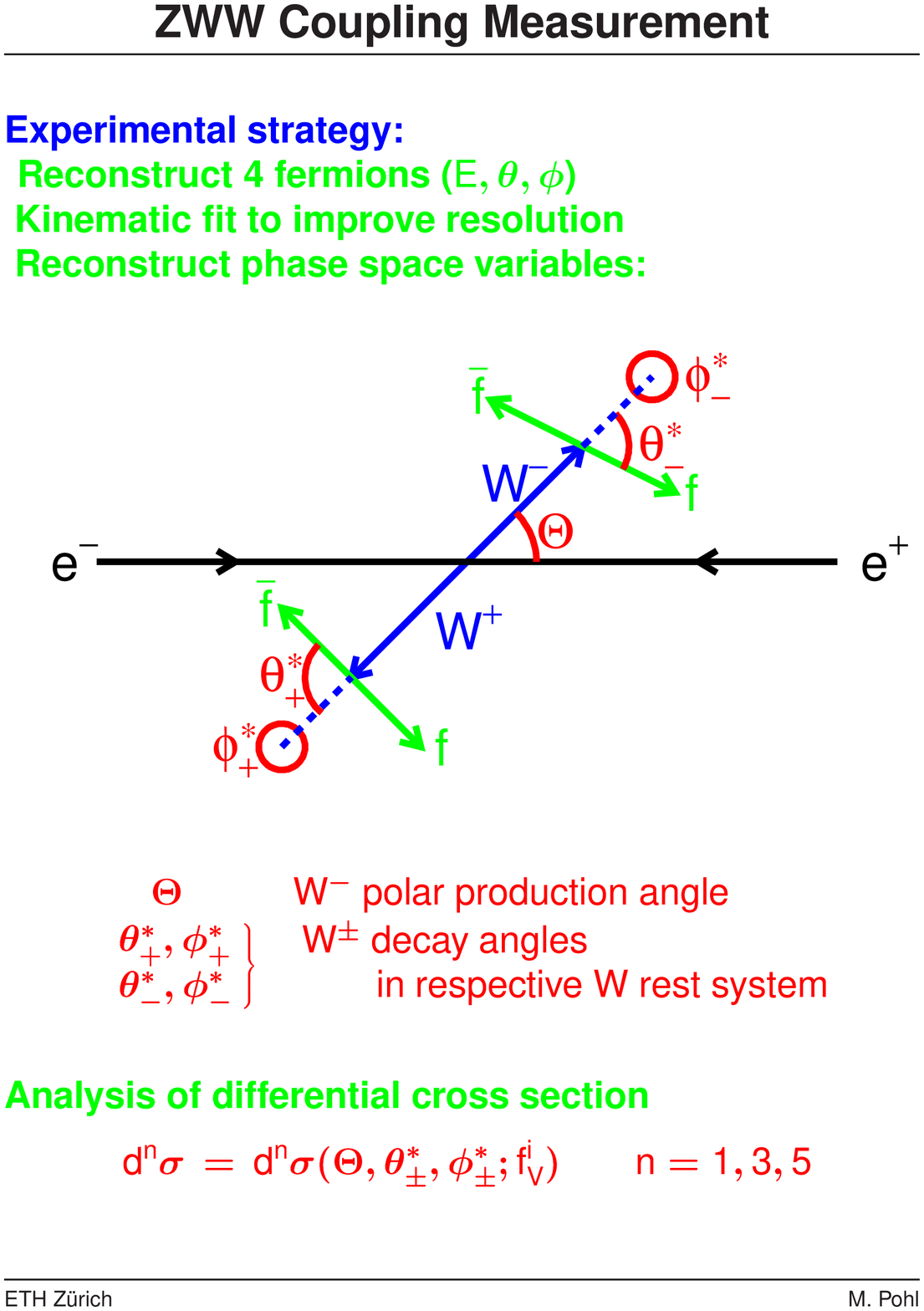}
  \end{center}
\label{fig:tgcang}
\caption{Production and decay angles in a WW event}
\end{figure}

In particular, the study of these couplings is performed with a combined fit
to the total
WW production cross section as well as the distribution of the production and
decay angle of each W boson (see figure \ref{fig:tgcang}).
\par
If no jet charge algorithm is used, the W production angle can only be 
unambiguously determined in semileptonic events, where the charge of the
lepton reflects the charge of the parent W. In the fully leptonic case,
all production and decay angles can be determined with a two-fold ambiguity
using kinematic criteria if initial state radiation and W width are neglected
\cite{lnlnreco}. In the four-jet channel the ambiguity on the angles can
only be solved using the jet charge. Algorithms based on the Feynman-Fields
approach \cite{feynfields} have correct charge identification probability 
around 70\%.\par
Detector effects in angle resolution and charge confusion are accounted for
using reweighting algorithms similar to those used for the measurement of
the W mass. An alternative approach is that based on Optimal Observables
\cite{optobs}. Instead of fitting the kinematic distributions, the 
differential cross section parametrized as a quartic function of the anomalous
couplings:
\[\frac{d\sigma}{d\Omega}=c_0(\Omega)+c_1(\Omega)\Psi+c_2(\Omega)\Psi^2\]
where $\Omega$ are the phase space variables, and $\Psi$ is one of the
couplings allowed to be different from the standard model, while the others 
are kept to zero. Assuming that all couplings are small, it is possible to
neglect the term $c_2$, and apart from this approximation the observable
\[O_1=\frac{c_1(\Omega)}{c_0(\Omega)}\]
contains all the information carried by the distributions $d\sigma/d\Omega$,
allowing the extraction of the couplings from a 1-dimensional fit.\par
\begin{figure}[p]
  \begin{center}
   \includegraphics[width=14cm,bb=45 215 525 735,clip]{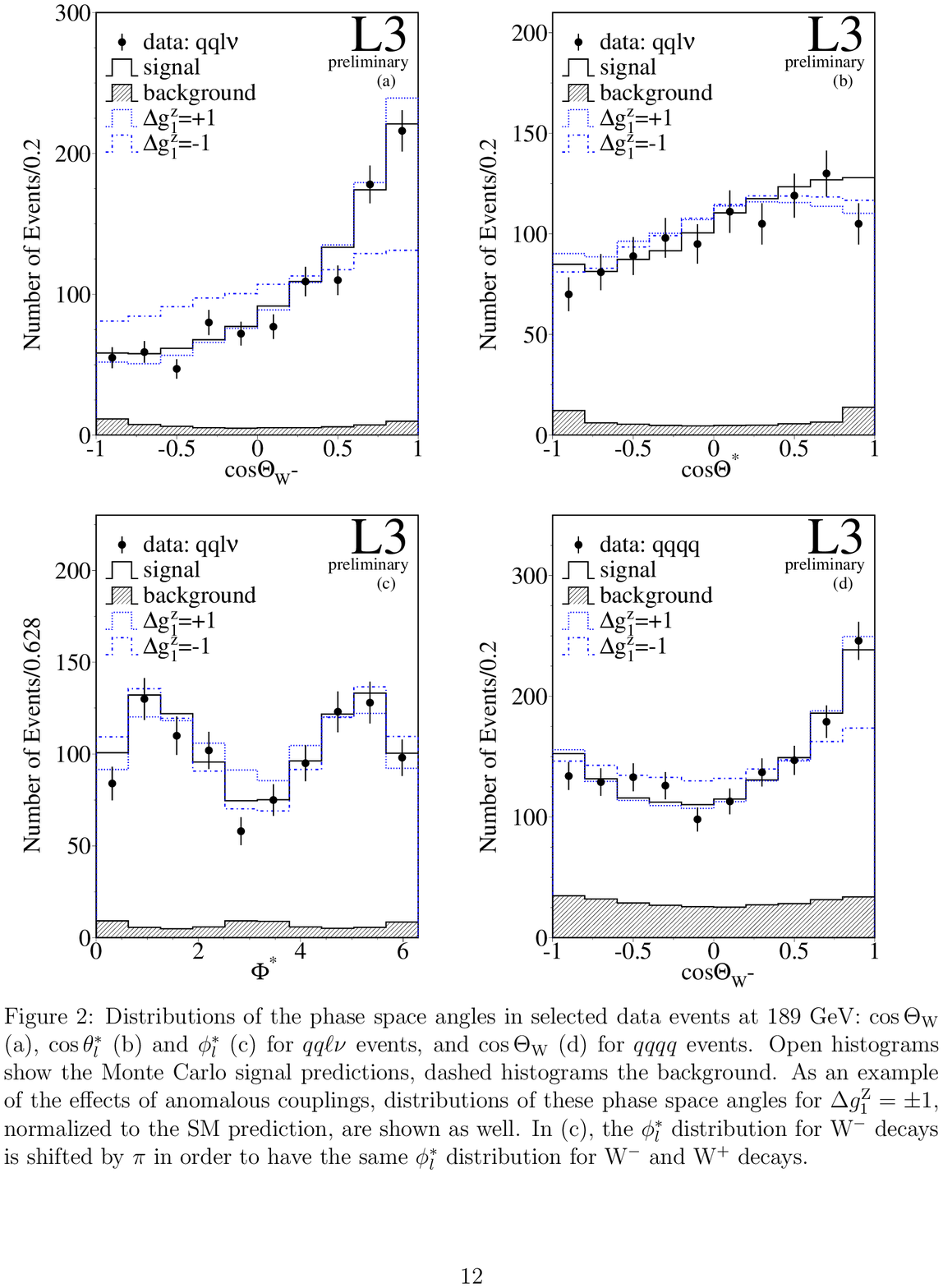}
  \end{center}
\label{fig:tgcvar}
\caption{Distributions of production and decay angles in L3 WW events at 189 GeV}
\end{figure}
\begin{figure}[p]
  \begin{center}
   \includegraphics[width=14cm,bb=17 153 560 715]{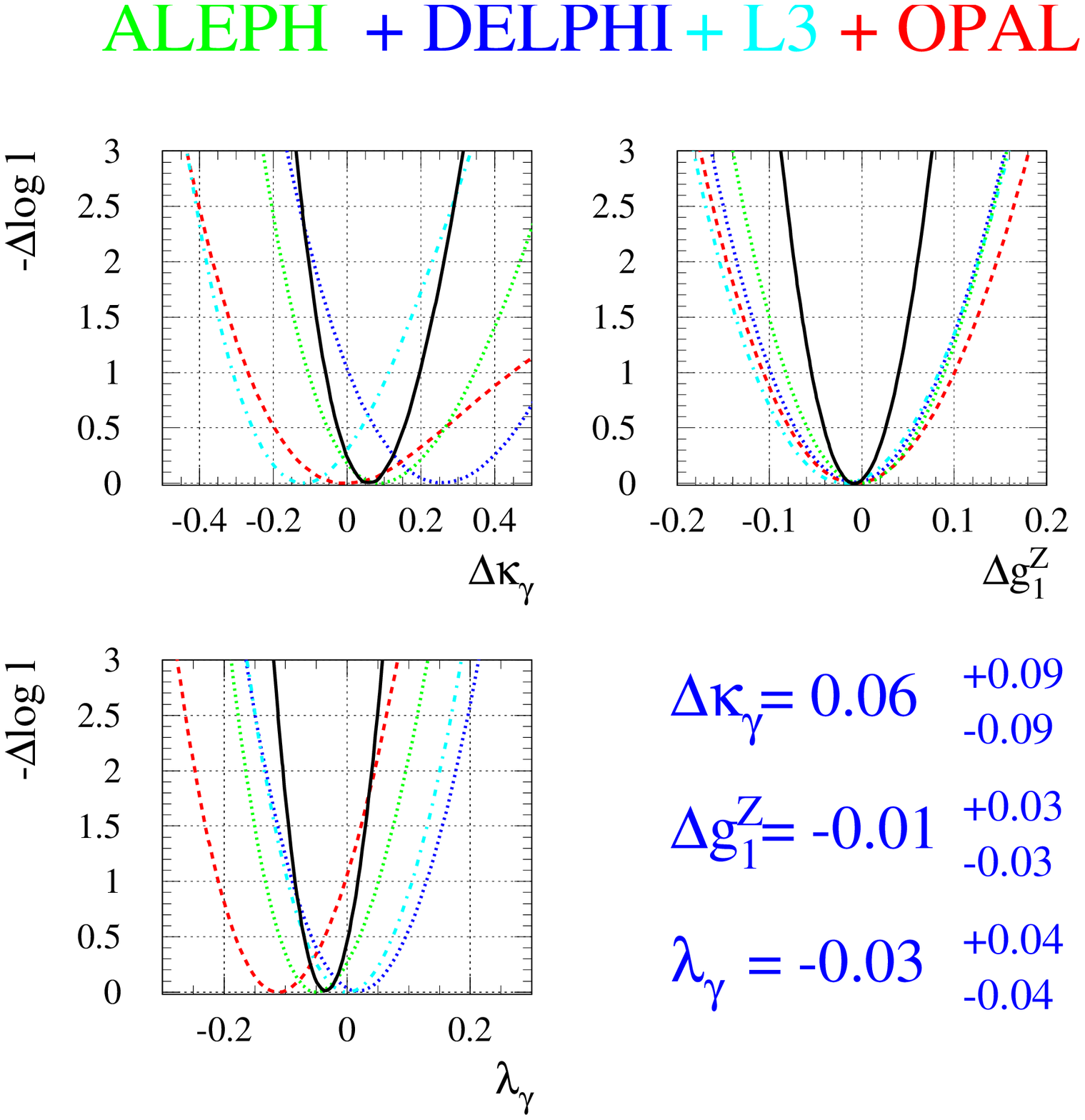}
  \end{center}
\label{fig:leptgc1}
\caption{Values of single coupling variables obtained combining the four LEP 
experiments, assuming the other variables set to the standard model value. The
four curves in each plot correspond to the results of the four LEP 
experiments.}

\end{figure}
\begin{figure}[p]
  \begin{center}
   \includegraphics[width=14cm,bb=0 120 550 700]{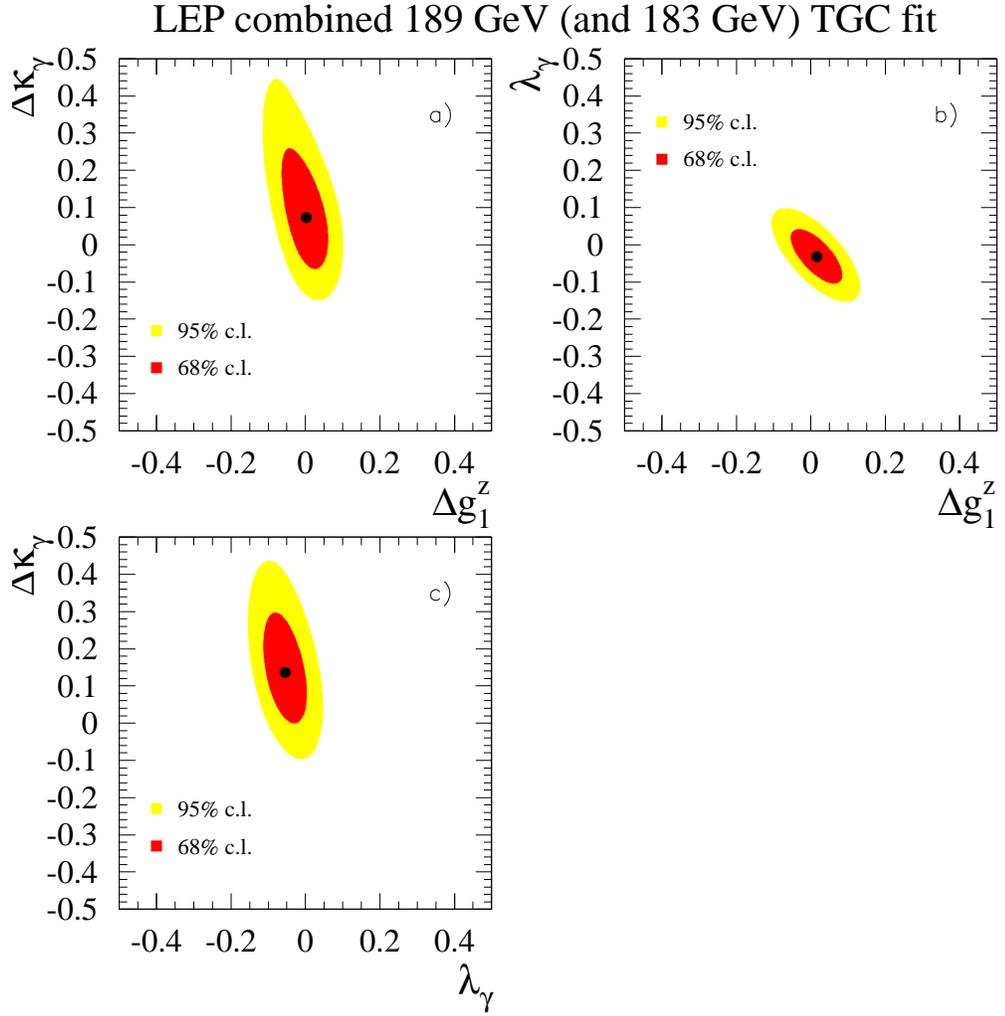}
  \end{center}
\label{fig:leptgc2}
\caption{Combined allowed contours for TGC variables when one is set to the 
standard model value and the other are left as free parameters of the fit.}
\end{figure}

\par
\begin{figure}[tbh]
  \begin{center}
   \includegraphics[width=7cm]{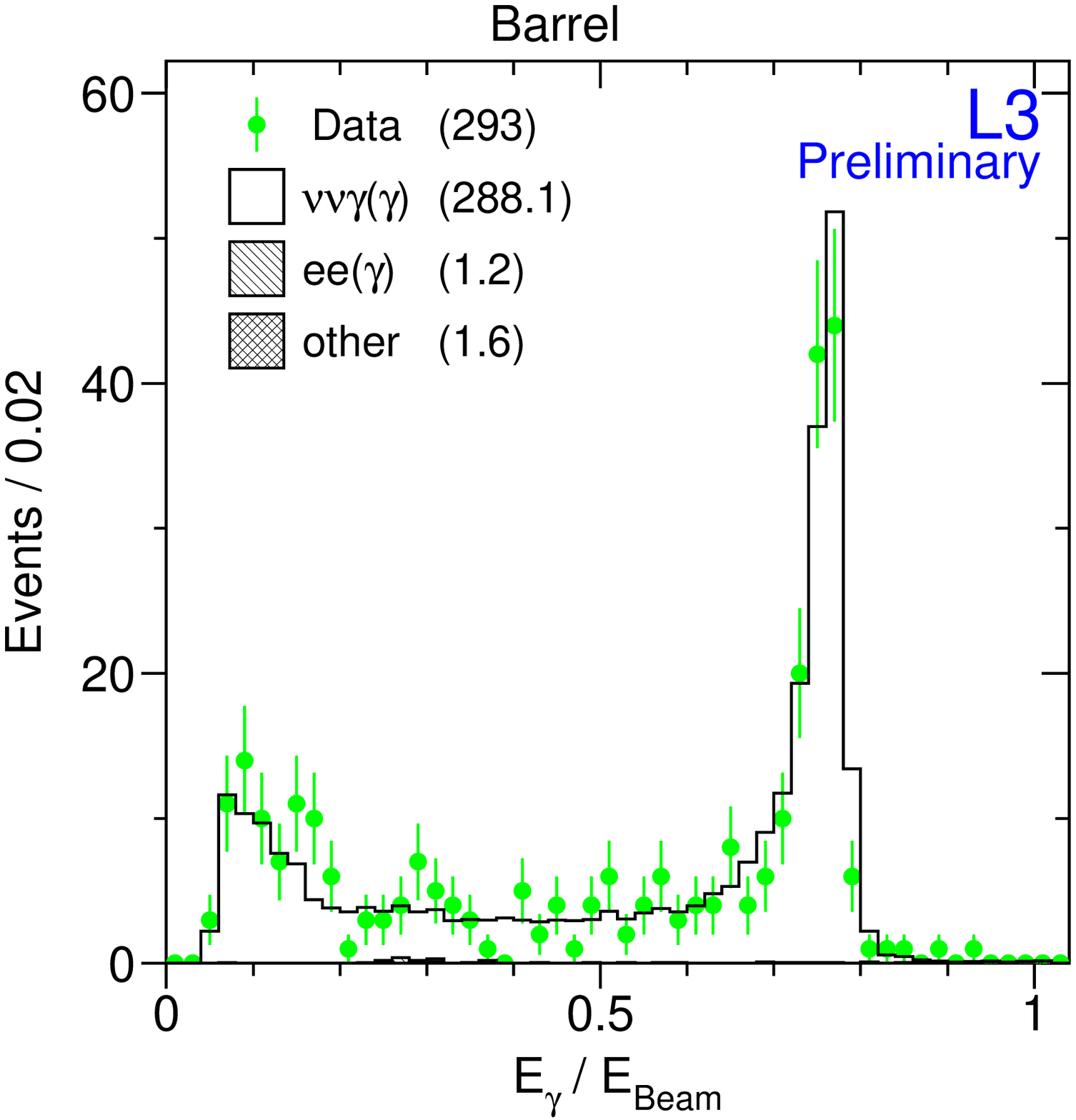}
  \end{center}
\label{fig:l3phot}
\caption{Energy spectrum of single photon 
events in L3 for 189 GeV data}
\end{figure}
\par
Diagrams involving the interaction of three vector bosons are not only present
in WW final state events, but also in other kind of processes, e.g. production
of single W and single photons (figure \ref{fig:l3phot}. With respect to WW 
production, these processes
are more sensitive to $\gamma WW$ couplings, $\Delta \kappa_\gamma$ and
$\Delta \lambda_\gamma$.\par
As discussed above, all analyses performed to study trilinear gauge couplings
are based on reconstructed distributions of W decay products (or photon).
The final sensitivity of the result strongly depends on the reconstruction 
performances for these quantities, therefore also in this case kinematic fits 
are widely used. The main sources of systematics are coming from the accuracy
of the MonteCarlo modeling of detector effects, as well as the effects in
fragmentation as well as final-state hadronic interactions, as already
discussed for the case of the W mass measurement.
\par
Final results for the couplings can be expressed in terms of one variable
constraining the others to the standard model values, or fitting two variables
at the same time. 
The LEP combined results following these two approaches are
shown in figures \ref{fig:leptgc1} and \ref{fig:leptgc2}. Results from the
single experiments can be found in the references \cite{tgcww}.
\par
\section{QCD effects}
Hadronic interactions occur between W decay products, and can lead to 
modification of the observed final states. In particular the understanding 
of these effects is very important for the W mass measurement, since they can 
lead to biases in the 4-jet channel far larger than the target accuracy for
this measurement. In particular, Bose-Einstein and color reconnection effects
will be discussed.\par
\subsection{Bose-Einstein effects}
Bose-Einstein correlations enhance the production of identical bosons close
in direction and momentum. These effects have already been observed in
nucleus-nucleus and hadron-hadron interactions\cite{beh}, as well as in 
hadronic Z decays at LEP1 \cite{belep1}.\par
In WW events, Bose-Einstein correlations can occur:\begin{itemize}
\item between bosons   within same jet 
\item between bosons   from same W 
\item between bosons   from different Ws 
\end{itemize}
Only the last case is important for W mass studies, since it generates 
distortions in the reconstructed invariant mass spectrum.\par
\begin{figure}[t]
 \begin{minipage}{.45\linewidth}
  \begin{center}
   \includegraphics[width=5cm,bb=63 206 510 725,clip]{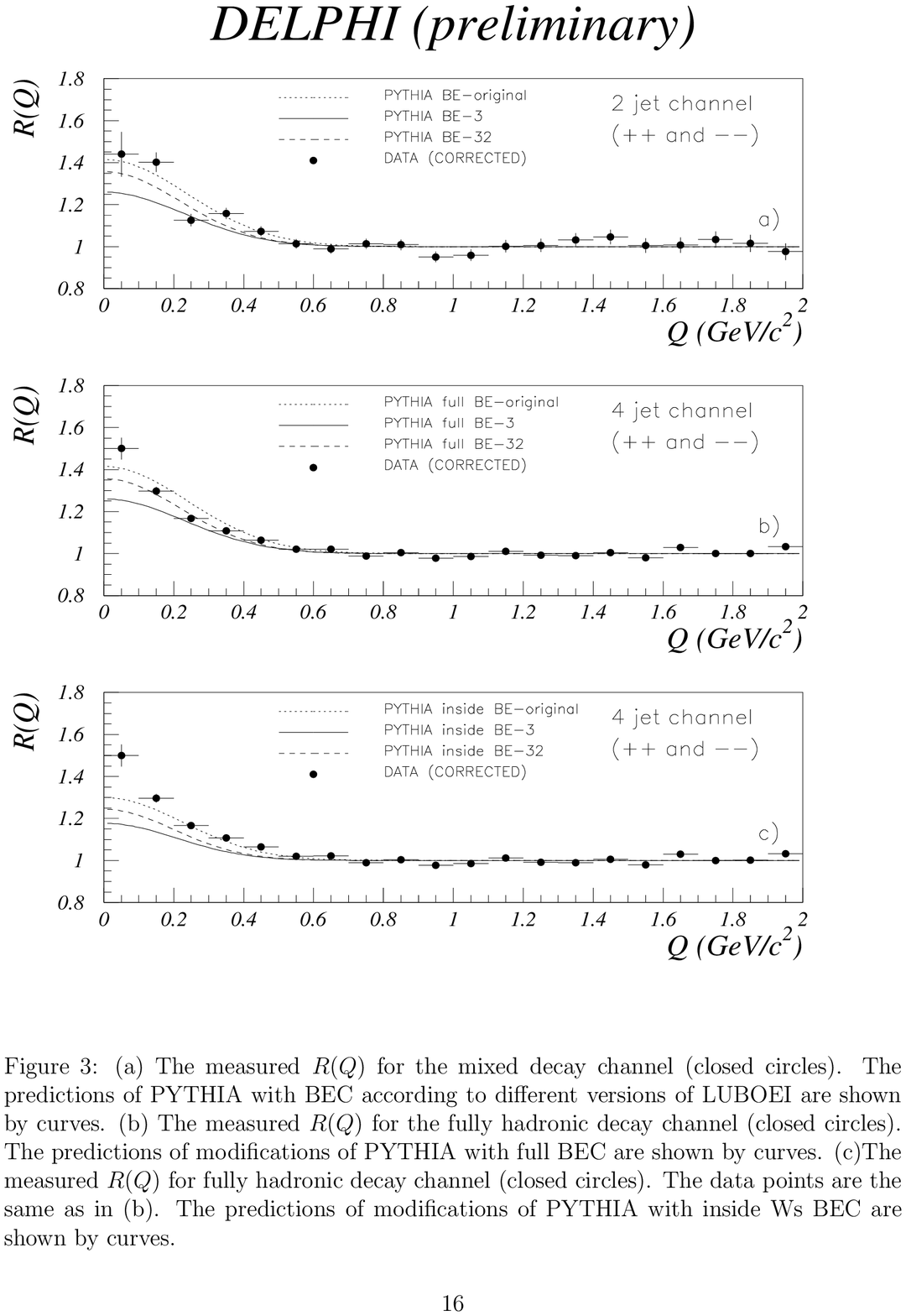}
  \caption{R(Q) distribution from DELPHI data at 183 and 189 GeV. (a): in 
semileptonic events. (b) in fully hadronic events, compared to a model with
full BEC. (c) same data as in (b), compared to a model with BEC only inside
the same jet. Data seem favoring models with full BEC.} \label{fig:delbe}
  \end{center}
 \end{minipage}
 \begin{minipage}{.45\linewidth}
  \begin{center}
   \includegraphics[width=6cm]{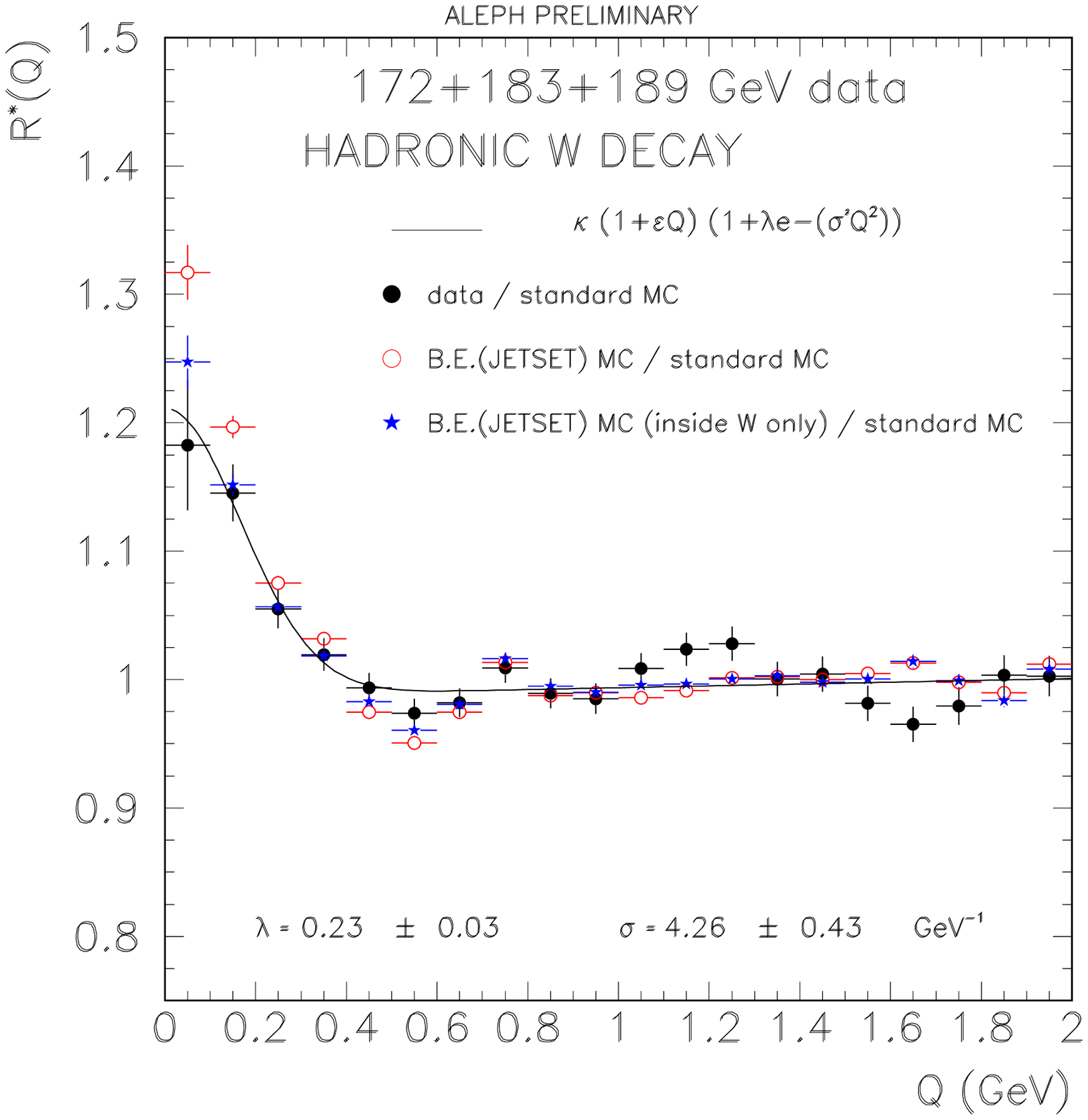}
  \caption{Data/MC ratio of R(Q) distributions in ALEPH. Data (full dots) are
in better agreement with MonteCarlo where BEC occur only inside the same W
(stars) than in models where they occur between different Ws (open dots)} 
\label{fig:albe}
  \end{center}
 \end{minipage}
\end{figure}
\par
To model this effect, the correlation function between two equal bosons is 
assumed to be Gaussian:
 
\[R(Q)=(1+\lambda e^{-Q^2R^2})\]  
where   $\lambda$   and   R   are the amplitude and the radius 
of the effect, and Q is the four-momentum difference between the two identical
bosons $Q^2=(p_1-p_2)^2$.\par
\par
From the experimental point of view, most of the effect shows up between
pions of the same charge inside hadronic jets, while pions of different charge
are not affected. Bose-Einstein correlations produces an enhancement in the
ratio of the Q distribution between same-sign ($\rho^{\pm\pm}$) and 
opposite-sign ($\rho{\pm\mp}$) pions:
\[R(Q)=\frac{\rho^{\pm\pm}(Q)}{\rho^{\pm\mp}(Q)}\]
The effect can either be seen from the R(Q) distribution itself (OPAL),
in the double ratio $R(Q)_{DATA}/R(Q)_{MC}$ (ALEPH, L3), or defining
\[R(Q)=\frac{\rho^{\pm\pm}(Q)}{\rho^{\pm\pm}_{MC}(Q)}\]
where the Monte Carlo sample has no Bose-Einstein correlations (DELPHI).\par
Delphi results are slightly in favor of the presence of Bose-Einstein
Correlations between particles from different Ws (figure \ref{fig:delbe}), 
while the results from ALEPH seem disfavoring (by 2.7 $\sigma$) such 
correlations (figure \ref{fig:albe}).\par
The preliminary values for the amplitude of the effect between pions coming 
from different W $\lambda^{diff W}$ are listed in table \ref{tab:bec}
\cite{beexp}. To be 
noticed that ALEPH and L3 results are preliminary, and the data sample used
are widely different. With the present available information the presence of
Bose-Einstein correlations between different Ws is still unclear.\par
\begin{table}[h]
\begin{center}
\begin{tabular}{|c|c|}\hline
Experiment&$\lambda^{diff W}$\\ \hline
ALEPH&$0.15\pm0.18$\\
DELPHI&$-0.20\pm 0.22$\\
L3&$0.75\pm 1.80$\\
OPAL&$0.22\pm 0.53$\\ \hline
\end{tabular}
\vspace{1cm}
\caption{\label{tab:bec}Fitted values for $\lambda$}
\end{center}
\end{table}
\par
\subsection{Colour reconnection}
String effects between jets coming from different Ws (but from partons with 
opposite colour) are another source of distortion of the mass distribution.
They can occur since the distance in space between the two W decay vertices
is of the order of 0.1 fm, while the hadronic scale is of the order of 1 fm.
\par
\begin{figure}[H]
  \begin{center}
   \includegraphics[width=6cm]{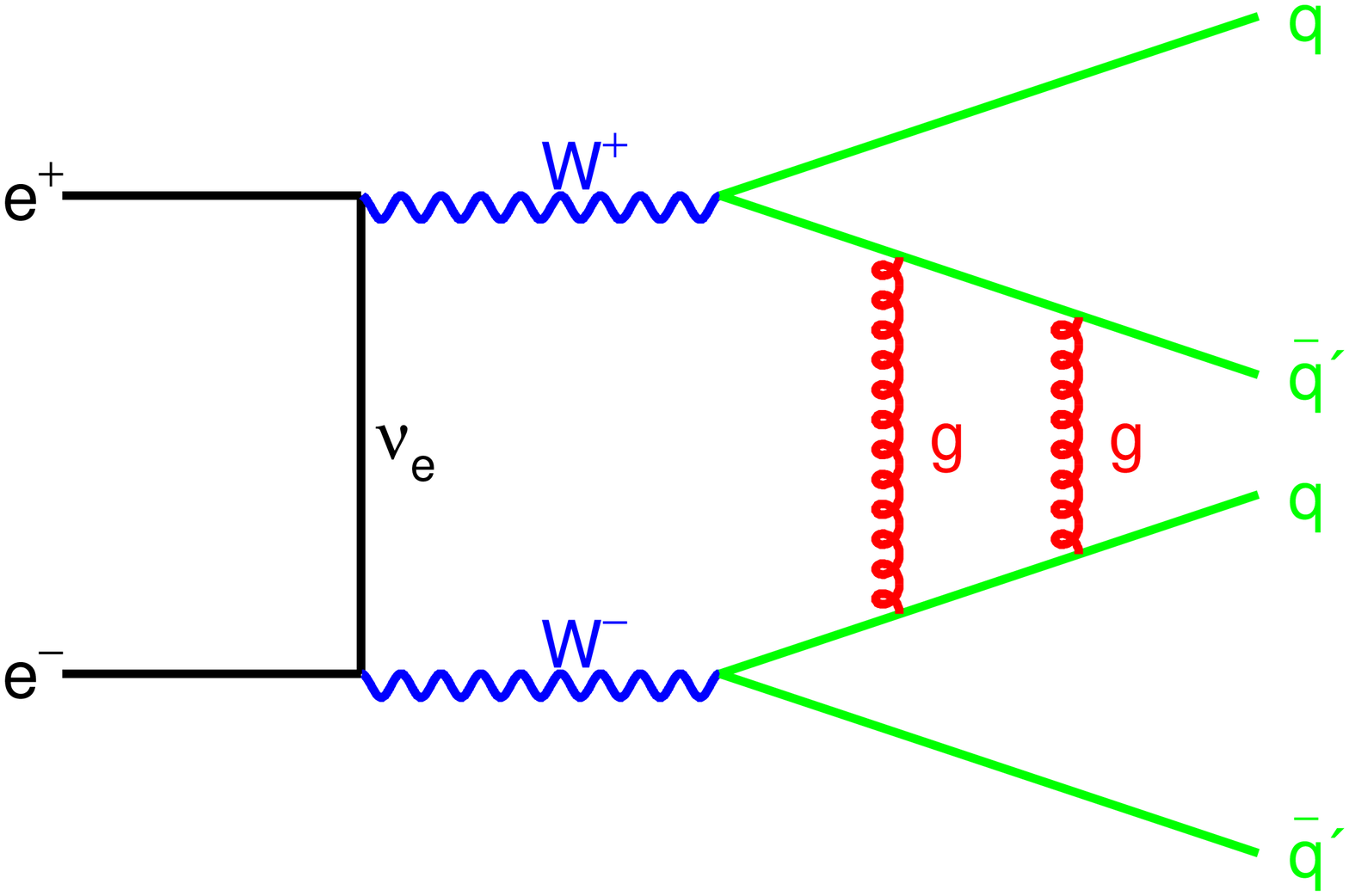}
  \end{center}
\end{figure}
\par

\begin{figure}[tb]
  \begin{center}
   \includegraphics[width=8cm]{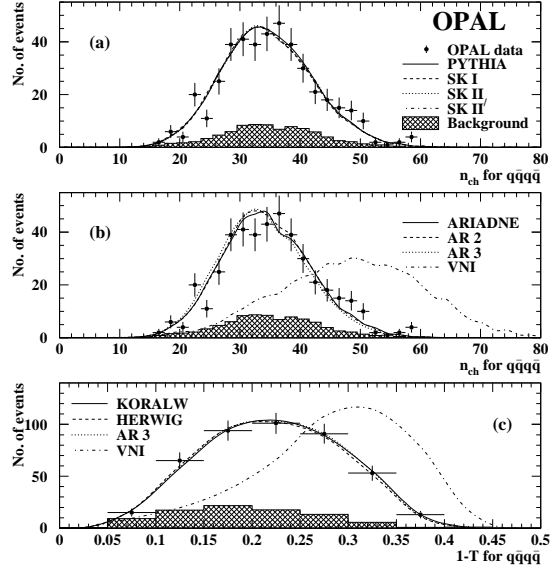}
  \end{center}
  \caption{Number of charged particles in 4-jet hadronic events compared to 
different color reconnection models. Apart from the VNI model, that fails to
reproduce the thrust distribution (plot c), the other models predict very
small shifts in the charged multiplicity, and the present data are not able to 
discriminate among them.}\label{fig:opalcr}
\end{figure}

Perturbative contributions are expected to be small,\cite{cr2} but the non perturbative 
part can lead to mass shifts of the order of some hundreds MeV, depending on
the model.
The experimental study of these effects exploits the fact that in addition 
to W mass shifts, colour reconnections produce modifications in the topology
of the events.
The most important observable used to discriminate among the different models 
is the charged multiplicity in four-jet events (often expressed as difference 
or ratio between charged particle production in $qqqq$ and $2\times qql\nu$ 
events, which are not affected by CR)\cite{cr1}. Typically,
models predicting shifts in the W mass of the order of several hundred MeV
also predict a difference in
the number of charged particles between semileptonic and hadronic events of
about 10\%, while for models leading to smaller mass shifts this
difference is few per cent. As shown if figure \ref{fig:opalcr}, it is very
difficult with present data to discriminate among the models implemented
in the event generators \cite{crall}, since
the predicted shift is similar and quite small.\par
\par
Table \ref{tab:cr} shows the difference (the ratio for DELPHI) between then
charged multiplicity of 4-jet events and twice the one of semileptonic events
\cite{crexp}.
Present errors on this quantity are still too large to be compared to the
existing models.\par
\begin{table}[h]
\begin{center}
\begin{tabular}{|c|c|}\hline
Experiment&$\Delta<n_{ch}>$\\ \hline
ALEPH&$0.47\pm0.44\pm0.26$\\
L3&$-.0\pm0.8\pm0.5$\\
OPAL&$0.7\pm0.8\pm0.6$\\ \hline
&$<N_{ch}^{qqqq}>/2<N_{ch}^{qq}>$\\ \hline
DELPHI&$0.977\pm0.017\pm0.027$\\ \hline
\end{tabular}
\vspace{1cm}
\caption{\label{tab:cr}Difference (ratio for DELPHI) in charged multiplicity
between four-jet and two-jet events. Results from L3 and OPAL do not include
189 GeV data. Only OPAL results are final.}
\end{center}
\end{table}
\par
Other observables for studying color reconnection studies are:\begin{itemize}
\item charged multiplicity in the low momentum region
\item track characteristic distributions (momentum, rapidity, $p_t$,...)
\item event shape (thrust,...)
\item heavy hadron multiplicity (K,p with $0.2 GeV< p < 1.4$ GeV)
\end{itemize}
Also for these observables no clear indications for color reconnection can be 
derived.
\par
\section{Charm production in W decays}
For real W bosons the production of b quarks is either forbidden by energy 
conservation (as in the case $W\to tb$) or strongly Cabibbo-suppressed
(as in the case of the decay $W\to cb$). For this reason, charm is the 
heaviest quark largely produced in W decays.
Using its heavy-quark characteristics in an almost b-free environment, it is 
possible to measure the charm production branching ratio in W decays.
In the standard model this value is precisely determined by the unitarity
of the CKM matrix:
\[\frac{|V_{cd}|^2+|V_{cs}|^2+|V_{cb}|^2}{|V_{cd}|^2+|V_{cs}|^2+|V_{cb}|^2+|V_{ud}|^2+|V_{us}|^2+|V_{ub}|^2}\]
\par
\begin{figure}[tb]
  \begin{center}
   \includegraphics[width=8cm]{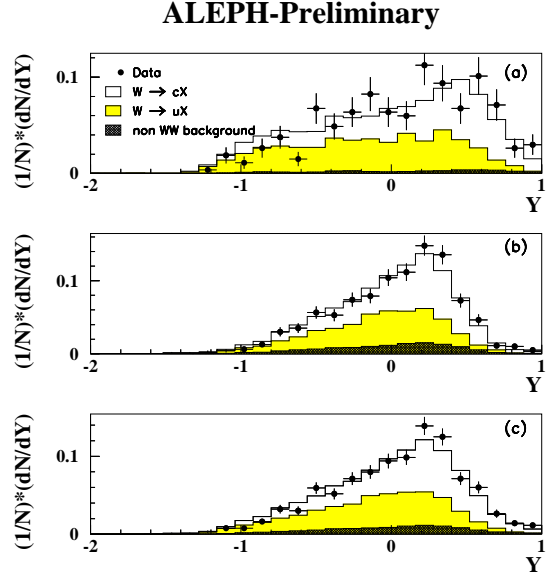}
  \end{center}
  \caption{Distribution of the Fisher Discriminant for (a) semileptonic
events (b) fully-hadronic events (c) the sum of the two classes}
\label{fig:alephfd}
\end{figure}
\par
For this reason, a measurement of the charm fraction in W decays is a direct 
test of the unitarity of the CKM matrix.
Furthermore, using the precise determinations of the other elements, it is
possible to convert this measurement into a determination of   $|V_{cs}|$.
\par
This approach to the determination of this matrix element is less precise 
than the derivation from the hadronic branching ratio, but more direct.
For charm tagging, the three experiments performing this measurement use 
different experimental techniques:
\begin{itemize}
\item   ALEPH (172+183 GeV data)   use a neural network with 12 input 
variables or a Fisher discriminator (see figure \ref{fig:alephfd}). 
Variables used come from b-tagging, event 
shapes, exclusive decays etc. A combination of the two methods is used for the
final result.
\item   L3 (183 GeV data)   splits jets into four categories, depending whether
an inclusive lepton (  $e,\mu$ ), a   $D^*\to D^0\pi^\pm$   
or none of the above is found. Each category is then analyzed by a separate 
neural network.
\item   DELPHI (172 GeV data)   exploits its RICH detector for kaon 
identification, thus directly measuring $V_{cs}$ through an s-tag in addition
to a charm tag performed in a similar way as the other experiments.
\end{itemize}
 
The results obtained are summarized in table \ref{tab:vcs}\cite{vcsexp}.
\begin{table}[H]
\begin{center}
\begin{tabular}{|c|c|}\hline
&$|V_{cs}|$\\ \hline
$D\to Kl\nu$(PDG)&$1.04\pm 0.16$\\ \hline
ALEPH&$1.00\pm0.10\pm0.06$\\
L3&$0.98\pm0.22\pm0.08$\\
DELPHI&$0.91\pm0.14\pm0.05$\\ \hline
LEP direct&$0.96\pm0.09$\\ \hline
LEP BR(W$\to$qq)&$1.03\pm0.04$\\ \hline
CKM unitarity&$0.9745\pm0.0005$\\ \hline
\end{tabular}
\vspace{1cm}
\caption{\label{tab:vcs}Determinations of $|V_{cs}|$}
\end{center}
\end{table}

\par
\section{Conclusions}
After three years of data taking above the WW threshold, W physics at LEP has
reached the realm of precision measurements. Data have been collected at
161, 172, 183 and 189 GeV of center of mass, with increasing integrated 
luminosity. The selection of WW events provides measurements of the WW
total cross section and of the branching ratios of all decay channels. 
All these measurement are in agreement with the predictions from the standard
electroweak theory. A sector where deviation from the standard theory could be
expected is the study of the trilinear gauge coupling. Fits to the cross 
section and to the kinematics of the events so far are in agreement with the
expectations, so limits on the presence of anomalous couplings are established.
Charm production in W decays provides a direct determination of $|V_{cs}|$,
while an indirect measurement can be derived from the hadronic branching
fraction. The most important measurement in W physics at LEP is the W mass.
After a derivation from the threshold cross section in the first run, the mass
is now measured using the direct reconstruction of the invariant mass of the
W decay products. The present combined result $M_W=80.368\pm0.065$ is as 
precise as the one obtained from hadronic machines, and its accuracy will 
further improve in the next years. The question is still open on whether the
accuracy on the W mass will be finally dominated by systematic uncertainties.
Some systematic errors are likely to considerably improve with more study, but 
the large uncertainties associated with QCD interactions in the four jet 
channel may be not so easy to reduce, and limit the final accuracy of the
result.\par
LEP2 will run for two more years, aiming to reach the design integrated
luminosity of 500 $pb^{-1}$ and trying to reach the centre-of-mass energy of
200 GeV. W physics will fully profit from
more data and from more energy points, and the results from the four LEP
experiments will increase their sensitivity both in the measurement of
the parameters of the standard theory and in the search for new phenomena.
\par

\newpage
\section{Acknowledgements}
I would like to thank M. Pohl for having introduced me to the field of W 
physics.
Many thanks also to M.Grunewald, L.Malgeri and A.Tonazzo and  for 
careful reading of the manuscript and A.Rubbia for the encouragement.

\nonumsection{References}

\end{document}